\documentclass[10pt,aps,prl,twocolumn,superscriptaddress]{revtex4}
\usepackage[linktocpage=true, colorlinks=true, urlcolor=blue, linkcolor=blue, citecolor=blue]{hyperref}
\usepackage{graphicx}
\usepackage{amsmath,amssymb}
\usepackage{color}
\usepackage{url}
\usepackage{hyperref}
\usepackage{cleveref}
\usepackage{xfrac}
\usepackage{upgreek}

\graphicspath{{./Images/}}

\begin{document}

\title{Waveform analysis of human retinal and choroidal blood flow with laser Doppler holography}

\author{L\'eo Puyo}
\affiliation{Corresponding author: gl.puyo@gmail.com}
\affiliation{Institut Langevin. Centre National de la Recherche Scientifique (CNRS). Paris Sciences \& Lettres (PSL University). \'Ecole Sup\'erieure de Physique et de Chimie Industrielles (ESPCI Paris) - 1 rue Jussieu. 75005 Paris France}

\author{Michel Paques}
\affiliation{Centre Hospitalier National d'Ophtalmologie des Quinze-Vingts, INSERM-DHOS CIC 1423. 28 rue de Charenton, 75012 Paris France}
\affiliation{Institut de la Vision-Sorbonne Universit\'es. 17 rue Moreau, 75012 Paris France}

\author{Mathias Fink}
\affiliation{Institut Langevin. Centre National de la Recherche Scientifique (CNRS). Paris Sciences \& Lettres (PSL University). \'Ecole Sup\'erieure de Physique et de Chimie Industrielles (ESPCI Paris) - 1 rue Jussieu. 75005 Paris France}

\author{Jos\'e-Alain Sahel}
\affiliation{Centre Hospitalier National d'Ophtalmologie des Quinze-Vingts, INSERM-DHOS CIC 1423. 28 rue de Charenton, 75012 Paris France}
\affiliation{Institut de la Vision-Sorbonne Universit\'es. 17 rue Moreau, 75012 Paris France}

\author{Michael Atlan}
\affiliation{Institut Langevin. Centre National de la Recherche Scientifique (CNRS). Paris Sciences \& Lettres (PSL University). \'Ecole Sup\'erieure de Physique et de Chimie Industrielles (ESPCI Paris) - 1 rue Jussieu. 75005 Paris France}

\date{\today}

\begin{abstract}
Laser Doppler holography was introduced as a full-field imaging technique to measure blood flow in the retina and choroid with an as yet unrivaled temporal resolution. We here investigate separating the different contributions to the power Doppler signal in order to isolate the flow waveforms of vessels in the posterior pole of the human eye. Distinct flow behaviors are found in retinal arteries and veins with seemingly interrelated waveforms. We demonstrate a full field mapping of the local resistivity index, and the possibility to perform unambiguous identification of retinal arteries and veins on the basis of their systolodiastolic variations. Finally we investigate the arterial flow waveforms in the retina and choroid and find synchronous and similar waveforms, although with a lower pulsatility in choroidal vessels. This work demonstrates the potential held by laser Doppler holography to study ocular hemodynamics in healthy and diseased eyes.
\end{abstract}

\maketitle

\section{Introduction}
The ocular circulation is of major interest for the study of increasingly prevalent diseases such as diabetic retinopathy, glaucoma, and hypertension. Other fundus diseases such as retinal vein occlusion and central serous chorioretinopathy are even more directly linked with the fundus perfusion. For the purposes of screening these diseases early on, understanding their pathophysiologies, and monitoring the efficiency of administrated treatments, measuring blood flow in the retina and choroid is of great importance. In the last decade, the irruption and development of optical coherence tomography (OCT) and subsequently of OCT-angiography (OCT-A) has allowed for a comprehensive study of the relationship between diseases affecting the eye and the fundus vasculature anatomy by its ability to resolve the retinal microvasculature~\cite{Kashani2017, ChuaChin2019}, and the thickness of the vascular choroidal compartment~\cite{Mrejen2013}. Another approach to investigate the ocular circulation is the study of the blood flow dynamics, which requires a temporal resolution currently not in the reach of present day OCT-A systems~\cite{Rosenfeld2016}. First, blood flow monitoring is important to evidence retinal hypoperfusion inherent to many ocular diseases. For example laser speckle flowgraphy can measure abnormal perfusion in glaucoma~\cite{Shiga2016, Mursch2018}, and Doppler ultrasound is able to detect decrease of the ocular circulation~\cite{Bonnin2014}. Going further into the study of blood flow dynamics, a time-resolved analysis of blood flow changes contains a wealth of information as the arterial waveform profile is influenced by the arterial compliance and peripheral resistance~\cite{McVeigh2007, Avolio2009, Rosenbaum2016}. Several approaches are used to investigate the Doppler waveform profiles such as the study of the arterial peak systolic and end diastolic blood flow velocity~\cite{Brar1988, Nichols2001}, which can be combined to produce indices to characterize the pulsatility~\cite{Maulik2005}. Another approach is to derive mathematical metrics from the pulse waveform curve such as skew, blowout time, rising rate and falling rate~\cite{Tsuda2014, Luft2016, GuZhang2018}. A wave decomposition analysis of ultrasonic Doppler flow velocity waveforms has also proven able to identify microvascular hemodynamic abnormalities in the ophthalmic and carotid artery in cases of diabetic retinopathy~\cite{Plumb2011}. Another factor to take into consideration for the shape and amplitude of the arterial flow profile is the site of measurement of the waveform. As reflection flow and pressure waves superimpose with the forward waves, they modify the incident arterial flow profile on a predictable basis with ageing and diseases~\cite{Mcveigh2002, Laurent2006, Safar2018}. Reflection waves are rapidly dissipated when propagating, thus it would be of great interest to measure the flow waveforms as close as possible from the arteriolar branching as it is considered to be the main sites of wave reflections~\cite{Safar2007}. This flow waveform is all the more interesting as measurements of an impaired pulsatile arterial flow could provide more information regarding potential microvascular damages provoked by an overly high perfusion pressure.

For the purpose of measuring blood flow directly in the retina, optic nerve head (ONH), and choroid, laser Doppler flowmetry (LDF) makes use of the time profile characteristics of the Doppler power spectrum density (DPSD)~\cite{Bonner1990, Riva1994, Riva2010}. Determining the Doppler broadening induced by the distribution of velocities and scattering angles allows to measure blood flow parameters such as volume and velocity. A short-time Fourier transform method is carried out to analyze the self-interfering light backscattered by the fundus, and blood flow variations are measured with a temporal resolution sufficient to observe the DPSD changes transient to cardiac cycles. However, this technique is performed only at a single location, and although full-field alternatives have been proposed~\cite{SerovLasser2005}, they have not been implemented in the retina as they require more illumination than permissible. Scanning implementation with the aim of imaging a two dimensional (2-D) field of view face a dilemma between temporal resolution and the spatial extent of the investigated area. Thus so far flying spot and line-scanning LDF have not demonstrated hemodynamic measurements over a 2-D field of view~\cite{Michelson1996, Mujat2019}. Laser Doppler holography (LDH) is a digital holographic method that relies on measuring the DPSD from the interference between a Doppler broadened beam and a holographic reference beam, allowing to extract a blood flow contrast similar to LDF~\cite{MagnainCastelBoucneau2014, Pellizzari2016, Donnarumma2016}. The additional reference beam is a plane or spherical wave and provides two advantages: it allows to use a higher camera frame rate by simply increasing the reference beam power, and it opens the door to all sorts of numerical manipulation of the wave field, such as numerical refocusing and digital adaptive optics~\cite{Hillmann2016, Ginner2018}, or synthetic subaperture Doppler measurements~\cite{Spahr2018, Ginner2019, Brodoline2019}. This way it becomes possible to perform a pixel wise parallel estimation of the DPSD in the human retina, which leads to simultaneous measurements of Doppler broadening over the 2-D field of view. We have previously used LDH to make full-field blood flow measurements in the retina with a few milliseconds of temporal resolution~\cite{Puyo2018}. We have then shown that the depth of field of a few hundreds microns provided by LDH allows to encompass both the retinal and choroidal vessels in a single imaging plane, and LDH is able to reveal non-invasively the choroidal vasculature while providing quantitative Doppler frequency shift measurements~\cite{Puyo2019}. We have used this feature to show the large differences of blood flow existing between arteries and veins in the choroid, and proposed an arteriovenous differentiation of choroidal vessels based on a flow analysis. In this article, we more carefully examine the waveform profiles that can be measured in retinal arteries and veins after spatio-temporal filtering, and show that the flow behavior is very different according to the vessel type. We then demonstrate the use of indices to characterize power Doppler waveforms: 2-D  maps of the local vascular resistivity index (RI) and coefficient of variation (CV) are shown. These metrics can be calculated over a few cardiac cycles and provide clear discrimination of arterial and venous flow in the retina. Finally, we analyze the dynamic flow in the choroid and demonstrate that the spatially averaged baseline signal (dominant signal) in laser Doppler measurements is close to the choroidal arterial waveform.

\section{Methods} \label{section_Methods}

\subsection{Optical setup}
We use the laser Doppler holographic setup presented in~\cite{Puyo2018}. Briefly, it consists of a fiber Mach-Zehnder optical interferometer where the light source is a $45 \, \rm  mW$ and $785 \, \rm  nm$ single frequency laser diode (Newport SWL-7513-H-P). It is an external cavity laser diode that is spatially and temporally coherent, with a 200 kHz spectral linewidth which is favorable for laser speckle contrast measurements~\cite{Postnov2019}. The power of the laser beam incident at the cornea is $1.5 \, \rm  mW$ of constant exposure and covers the retina over a disk with a radius of 3 to 4 mm. This irradiation level is compliant with the exposure levels of the international standard for ophthalmic instruments ISO 15004-2:2007. Informed consent was obtained from the subjects, experimental procedures adhered to the tenets of the Declaration of Helsinki, and study authorization was sought from the appropriate local ethics review boards (CPP and ANSM, IDRCB number: 2019-A00942-5). The results shown were obtained from 7 eyes of 5 different subjects, with ages ranging from 24 to 33.

The wave backscattered by the retina is combined with the reference wave using a non-polarizing beamsplitter cube. The polarization of the reference wave is adjusted with a half-wave plate and a polarizer to optimize fringe contrast. The interferograms formed in the sensor plane are recorded using a CMOS camera (Ametek - Phantom V2511, quantum efficiency 40\%, 12-bit pixel depth, pixel size $28 \, \upmu \rm m$) with a frame rate $f_{S}$ of typically 60 or $75 \, \rm  kHz$  in a 512 $\times$ 512 format. The diffracted speckle pattern is numerically propagated to the retinal plane using the angular spectrum propagation method~\cite{Goodman2005}. The configuration is on-axis and the reconstruction distance is large enough so that the holographic twin image energy is spread over the reconstructed hologram and has no appreciable effect on the resulting image. The following data processing consists of measuring the local optical temporal fluctuations of the holograms cross-beating terms in order to make quantitative measurements from the DPSD.

\subsection{Doppler processing}
After numerical propagation of the recorded interferograms, the beat frequency spectrum of the reconstructed holograms is analyzed pixel-wise by short-time Fourier transform:
\begin{equation} \label{eq:eq_PSD}
S(x,y,t_{n},f) = \left| \int_{t_{n}}^{t_{n}+ t_{\rm win}} H(x,y,\tau) \exp{\left(-2 i\pi f \tau\right)} \, {\rm d}\tau \right|^2 
\end{equation}
Where $H(x,y,\tau)$ is the hologram complex amplitude at pixel indexes (x,y) and time $\tau$, and $t_{\rm win}$ is the integration time of the sliding window which determines the temporal resolution of the system. In order to optimize the trade-off between signal-to-noise ratio (SNR) and temporal resolution ($t_{\rm win}$), we typically use a short-time window with 512 images ($6.8 \, \rm  ms$ with a $75 \, \rm  kHz$ frame rate) and an overlap between two consecutive windows of 256 images.

The Doppler spectrum contains a lot of hemodynamic information that can be retrieved through the changes over time of amplitude and frequency of the spectrum. In the conventional data processing for optical~\cite{Bonner1990}, and ultrasound Doppler measurements~\cite{Evans1989, Mace2011}, the zeroth moment, first moment, and normalized first moment of the DPSD reveal local blood volume, blood flow, and blood velocity, respectively. The zeroth and first moment of the DPSD are here denoted $M_{0}(x,y,t_{n})$ and $M_{1}(x,y,t_{n})$, and are calculated in the following way:
\begin{equation}\label{eq:eq_Moments}
M_{0}(x,y,t_{n}) = \int S(x,y,t_{n},f) \;  df
\end{equation}
\begin{equation}\label{eq:eq_Moments}
M_{1}(x,y,t_{n}) = \int S(x,y,t_{n},f) \; f  \; df
\end{equation}

In the rest of the article, the zeroth moment is also referred to as "power Doppler"; it is computed as the area under the power spectrum curve so it represents the number of scatterers traveling at any speed above the frequency threshold. Consequently it depends on the quantity of moving red blood cells, i.e. the blood volume in a given pixel. The frequency content of the spectrum reflects the velocity of the sampled area; it is analyzed using the normalization of the first moment by the zeroth moment, which mathematically yields the mean Doppler frequency shift:
\begin{equation}\label{eq:eq_Moments}
f_{\rm mean}(x,y,t_{n}) = \frac{\int S(x,y,t_{n},f) \; f  \; df}{\int S(x,y,t_{n},f) \; df} = \frac{M_{1}(x,y,t_{n})}{M_{0}(x,y,t_{n})}
\end{equation}

We have empirically determined that for $f_{\rm mean}$ the best results were obtained when removing the first term of the Fourier spectrum, i.e. with a frequency range of [0.5 kHz - $f_{S}/2$]. The interferometric zero-order term is naturally removed by the spectrum analysis which gives more weight to higher frequencies. As for $M_{0}$, the upper integration boundary is also generally set to $f_{S}/2$, but a higher low cut-off frequency is used so as to remove bulk motion noise. This low cut-off frequency is typically set to $6 \, \rm  kHz$ to obtain a power Doppler image where smaller vessels can be revealed, or $10 \, \rm  kHz$ if the aim is to measure a waveform without any contribution from ocular movement in the signal (except for micro-saccades). The result of this high-pass filtering is that $M_{0}$ not only depends on blood volume but also on blood velocity as when the flow speed increases, a part of the spectrum reaches the frequency threshold and becomes integrated in $M_{0}$. This effect is analyzed in section~\ref{FrequencyRange}.

Finally $M_{0}(x,y,t_{n})$ and $M_{1}(x,y,t_{n})$ images are compensated for the non-uniform illumination of the retina and vignetting. To do so, each image is divided by its spatially filtered (blurred) self. A conservation of the image energy (calculated as the sum of all pixel intensity) is used so that each image has the same energy before and after division. Only the spatial distribution of energy is changed while the total energy is conserved.

\subsection{Indices to characterize the Doppler waveform}
Several indices quantitatively characterizing Doppler waveform profiles have been introduced so as to study circulatory parameters~\cite{Maulik2005}. In this approach, the vascular circulation is described as an alternative current circuit, where the blood flow in the arterial system is a pulsatile phenomenon driven by periodic myocardiac contractions, and whose waveform profile is shaped by both upstream and downstream circulatory parameters~\cite{Nichols1980, Toy1985}. The higher compliance at the level of the large elastic arteries helps buffering the intermittent inflow in order to convert it into a steadier flow. The end of the vascular tree also plays an important role by offering a vascular resistance to the incoming flow at the arteriolar level. The opposition to blood flow due to friction on the vessel walls encompasses parameters such as blood viscosity, vessel length, and total cross-sectional area of the arterial bed. In the analogy with electrical circuits, the arterial compliance plays a capacitive role and the downstream opposition to flow acts as a resistance. Therefore, the flow wave is shaped by fundamental properties of the vasculature, and analyzing the Doppler waveform is of great interest as a surrogate means to assess these properties. The indices used for the purpose of studying the hemodynamic modulation of the Doppler waveform have various discriminatory performances and abilities to reflect different circulatory parameters; their use has become standard in Doppler ultrasound~\cite{Tranquart2003, Maulik2005, Demene2014}. In this work, we used two of these metrics: the first one is known as the Pourcelot or resistivity index (RI)~\cite{Pourcelot1974}. For a given function $g$, the RI is calculated pixel-wise using the following formula:
\begin{equation}\label{eq:eq_Pourcelot}
RI_{g}(x,y) = \frac{g(x,y,t_{\rm syst}) - g(x,y,t_{\rm diast})}{g(x,y,t_{\rm syst})}
\end{equation}
Where $g(x,y,t)_{\rm syst}$ and $g(x,y,t)_{\rm diast}$ are the values of the function $g$ at the peak systolic and end diastolic time points, which are manually determined. In a compliant vasculature, the resisitivity index increases with the vascular resistance peripheral to the measurement site~\cite{Bude1999Relationship, Bude1999Downstream, Halpern1998}. However it is also subject to changes of arterial compliance as a reduced upstream arterial compliance produces an impaired pulsatile flow with abnormally high systolic flow and lower than normal diastolic flow. Therefore it has been suggested to name it impedance index instead~\cite{Bude1999Relationship, Polska2001}. The second metric we used is a simple estimation of the local coefficient of variation of the power Doppler temporal fluctuations, calculated as the local standard deviation normalized by the local average.
\begin{equation}\label{eq:eq_CoefficientVariation}
CV_{g}(x,y) = \frac{\sigma_{g(x,y)}}{\mu_{g(x,y)}}
\end{equation}
Where $\sigma_{g(x,y)}$ and $\mu_{g(x,y)}$ denote the standard deviation and average of temporal fluctuations calculated pixel-wise over a few cardiac cycles. With both the $RI$ and $CV$ metrics, the areas where the systolodiastolic variations are greater come out with greater values, whereas areas with lower variations have lower values.

\section{Results in the retina}
\subsection{Pulsatile flow in retinal arteries and veins}
\begin{figure}[t!]
\centering
\includegraphics[width = 1\linewidth]{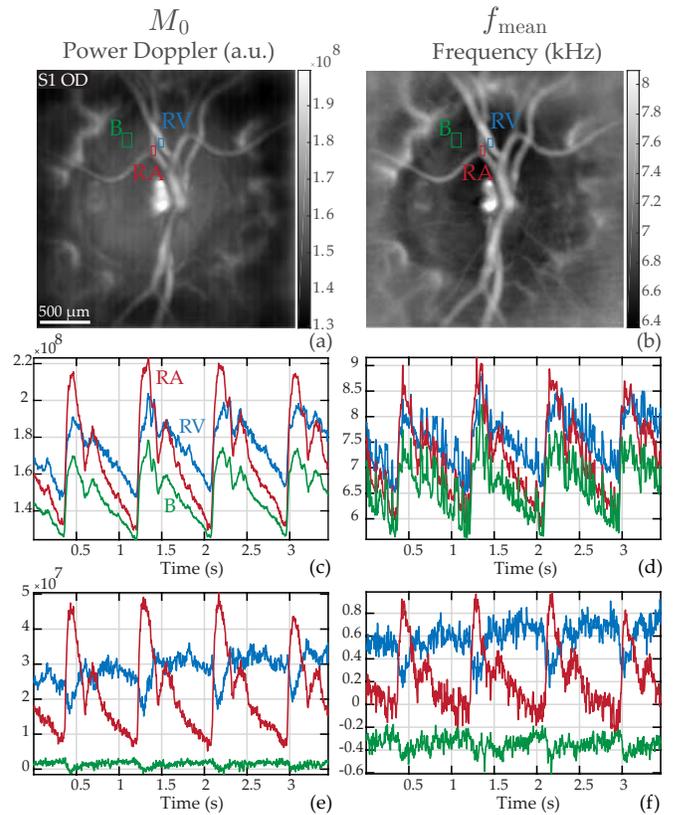}
\caption{Analysis of the dynamic changes in power Doppler signal and mean Doppler frequency shift, which supposedly yield volume and velocity, respectively. (a) and (b): power Doppler image $M_{0}(x,y)$ and mean frequency shift $f_{\rm mean}(x,y)$ temporally averaged; the red, blue, and green ROIs mark a retinal artery (RA), a retinal vein (RV), and the background (B), respectively. (c) and (d) temporal traces of $M_{0}$ and $f_{\rm mean}$ in the depicted ROIs. (e) and (f): temporal traces in the same ROIs when subtracting the spatially averaged signal over the entire image, i.e. $M_{0} - \langle M_{0} \rangle_{x,y}$, and $f_{\rm mean}- \langle f_{\rm mean} \rangle_{x,y}$. See \textcolor{blue}{\href{https://osapublishing.figshare.com/articles/media/Visualization_1/7988219}{Visualization 1}} and \textcolor{blue}{\href{https://osapublishing.figshare.com/articles/media/Visualization_2/7988228}{Visualization 2}} for the juxtaposed movies of $M_{0}$ and $f_{\rm mean}$ corrected and not-corrected from the spatial average. Removing the spatial average allows to reveal the retinal flow waveforms.
}
\label{fig_2_HighFrequencyShifts}
\end{figure}

In Fig.~\ref{fig_2_HighFrequencyShifts}, we analyze the temporal changes in power Doppler $M_{0}(x,y,t_{n})$ and mean Doppler frequency shift $f_{\rm mean}(x,y,t_{n})$ in different regions of interest (ROIs). The LDH acquisition was performed close to the optic nerve head, at 75 kHz, and $M_{0}$ was integrated over the frequency range [10-37 kHz] to fulfill Nyquist requirement, while we used the frequency range [0.5-37 kHz] for the calculation of $f_{\rm mean}$. In Fig.~\ref{fig_2_HighFrequencyShifts}(a) and (b), we show the power Doppler, and mean Doppler frequency shift images averaged over time. In Fig.~\ref{fig_2_HighFrequencyShifts}(a-b) are shown the ROIs used to measure the local pulsatile flow; the red, blue, and green boxes mark an artery, a vein, and an area in the background devoid of any large vessels. In Fig.~\ref{fig_2_HighFrequencyShifts}(c) and (d), we have plotted the local dynamic power Doppler and mean frequency shift variations in the depicted ROIs and found very similar variations in all areas. So, in Fig.~\ref{fig_2_HighFrequencyShifts}(e) and (f), the baseline signal (spatially averaged  over the entire field of view, also referred to as dominant signal), is subtracted from the same measurements, i.e. we measure $M_{0} - \langle M_{0} \rangle_{x,y}$, and $f_{\rm mean}- \langle f_{\rm mean} \rangle_{x,y}$. The difference between the raw movie and the movie corrected from the spatial average are shown in \textcolor{blue}{\href{https://osapublishing.figshare.com/articles/media/Visualization_1/7988219}{Visualization 1}} and \textcolor{blue}{\href{https://osapublishing.figshare.com/articles/media/Visualization_2/7988228}{Visualization 2}}, for $M_{0}$ and $f_{\rm mean}$.

The power Doppler and mean Doppler frequency shift images are similar but some difference of contrast can be noticed in certain features, especially the optic disc appears darker with the mean frequency image than with the power Doppler image. As mentioned in section~\ref{section_Methods}, the power Doppler supposedly reflects the local blood volume while the mean frequency shift is related to the local blood velocity; $f_{\rm mean}$ is measured in absolute frequency units (kHz) whereas $M_{0}$ is measured in arbitrary units (a.u.). A first observation on the signal dynamic changes that can be made from Fig.~\ref{fig_2_HighFrequencyShifts}(c) and (d), is that the raw measurements performed with LDH are very correlated between all the ROIs. For both the power Doppler and mean frequency shift, all measured waveforms have very similar profiles. The systolic and diastolic times are very well visible, and the dicrotic notch is also noticable, in the artery, vein and background. When subtracting $\langle M_{0} \rangle_{x,y}$ and $\langle f_{\rm mean} \rangle_{x,y}$ from the dynamic signal as shown in Fig.~\ref{fig_2_HighFrequencyShifts}(e) and (f), significantly different behaviors can be observed accordingly with the probed feature. Indeed, with this process, the signal in the retinal artery shows considerably larger systolodiastolic variations whereas the signal in the retinal vein appears to be more constant, almost cycloidal; the dicrotic notch becomes clearly visible only in the artery and not in the vein. In the arterial profile, the systolic upward slope is very steep while the diastolic downward slope is gentle. Finally, for both the power Doppler and mean frequency shift, the arterial systolic peak maximum coincides with the minimum of flow in the retinal vein.

Subtracting the baseline signal allows to reveal flow behaviors specific to the probed features. This operations aims at removing the contribution of underneath features such as multiple scattering from the choroid. The spatially averaged signal subtracted from the dynamic data also has an effect on the images by adding an offset value to the power and mean frequency shift maps. A linear correction should be applied to derive the actual physical units~\cite{Bonner1990}. The variations over time measured in terms of power Doppler and mean frequency shift are quite similar, although $f_{\rm mean}$ is more noisy. This is because unlike for $M_{0}$, the low frequency range [0.5-10 kHz] is used in the calculations of $f_{\rm mean}$, and this frequency range carries a contribution from global eye movements. However the time-averaged mean frequency image seems of slightly better quality than the power Doppler image. In the rest of the article we represent images mostly with the mean frequency shift metric, and temporal line plots of the hemodynamic with the power Doppler metric.

\subsection{Dependence of the waveform profile upon the frequency range} \label{FrequencyRange}

\begin{figure}[t!]
\centering
\includegraphics[width = 1\linewidth]{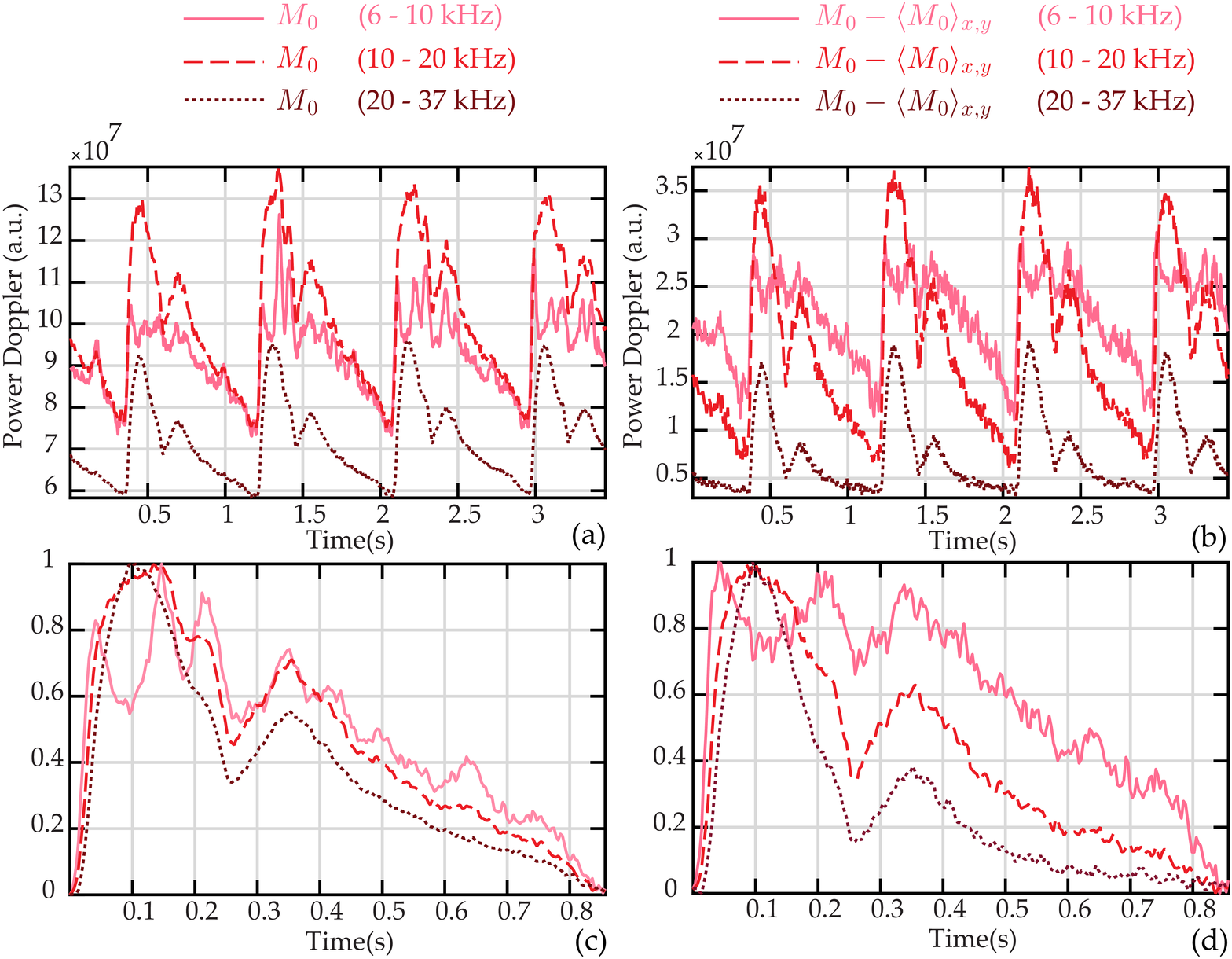}
\caption{Dependence of the retinal arterial waveform profile upon the Doppler frequency range. All curves are obtained from the same LDH measurements, only with different processing. (a) and (b): dynamic raw power Doppler and power Doppler corrected from the spatial average for three frequency ranges: 6-10 kHz in the pink solid line, 10-20 kHz in the red dashed line, and 20-37 kHz in the deep red dotted line. (c) and (d): waveform profile averaged over three cardiac cycle with variations normalized between 0 and 1. The waveform profile calculated with the frequency band 6-10 kHz shows a pulsatile signal tainted by bulk motion (attributed by unexpected oscillations); the normalized arterial waveform calculated with the very high frequency range appears steeper due to the velocity thresholding effect.
}
\label{fig_2_SeveralFrequencyRanges}
\end{figure}

In Fig.~\ref{fig_2_SeveralFrequencyRanges}, we analyze the arterial waveform profile as a function of the frequency range used for the power Doppler calculation. Figure ~\ref{fig_2_SeveralFrequencyRanges}(a) and (b) show the dynamic raw power Doppler and power Doppler corrected from the spatial average in an artery for three frequency ranges: 6-10 kHz in the pink solid line, 10-20 kHz in the red dashed line, and 20-37 kHz in the deep red dotted line. In Fig.~\ref{fig_2_SeveralFrequencyRanges}(c) and (d), we show the corresponding waveform profiles averaged over three cardiac cycles with variation normalized between 0 and 1.

A first observation is that the waveform profile calculated with frequencies under 10 kHz resembles the pulse waveform, but with additional noise. This is presumably because this lowest frequency range contains a mix of spectral contributions from bulk motion and pulsatile blood flow. Thus even though a part of the investigated flow signal lies in the frequency range 6-10 kHz, it seems best not to use this frequency range because it is tainted by global eye movements. In order to obtain a waveform purely representative of pulsatile blood flow, in the rest of the article we work with a 10 kHz lower frequency threshold for the calculation of $M_{0}$. The second observation is that the very high Doppler frequency shifts normalized arterial waveform exhibits a steeper profile as visible from the difference between the 10-20 and 20-37 kHz profiles in Fig.~\ref{fig_2_SeveralFrequencyRanges}(d). This is the result of the high-pass frequency filtering which acts as a velocity threshold. This thresholding removes the contribution of the lower speed constant flow as the velocity limit is exceeded mostly during systole and dicrotic notch. This leads to a sharper peak during systole and dicrotic notch, which amplifies the pulsatile shape of the blood flow waveform.

\subsection{Phase relationship between the arterial and venous waveforms}
\begin{figure}[t!]
\centering
\includegraphics[width = 1\linewidth]{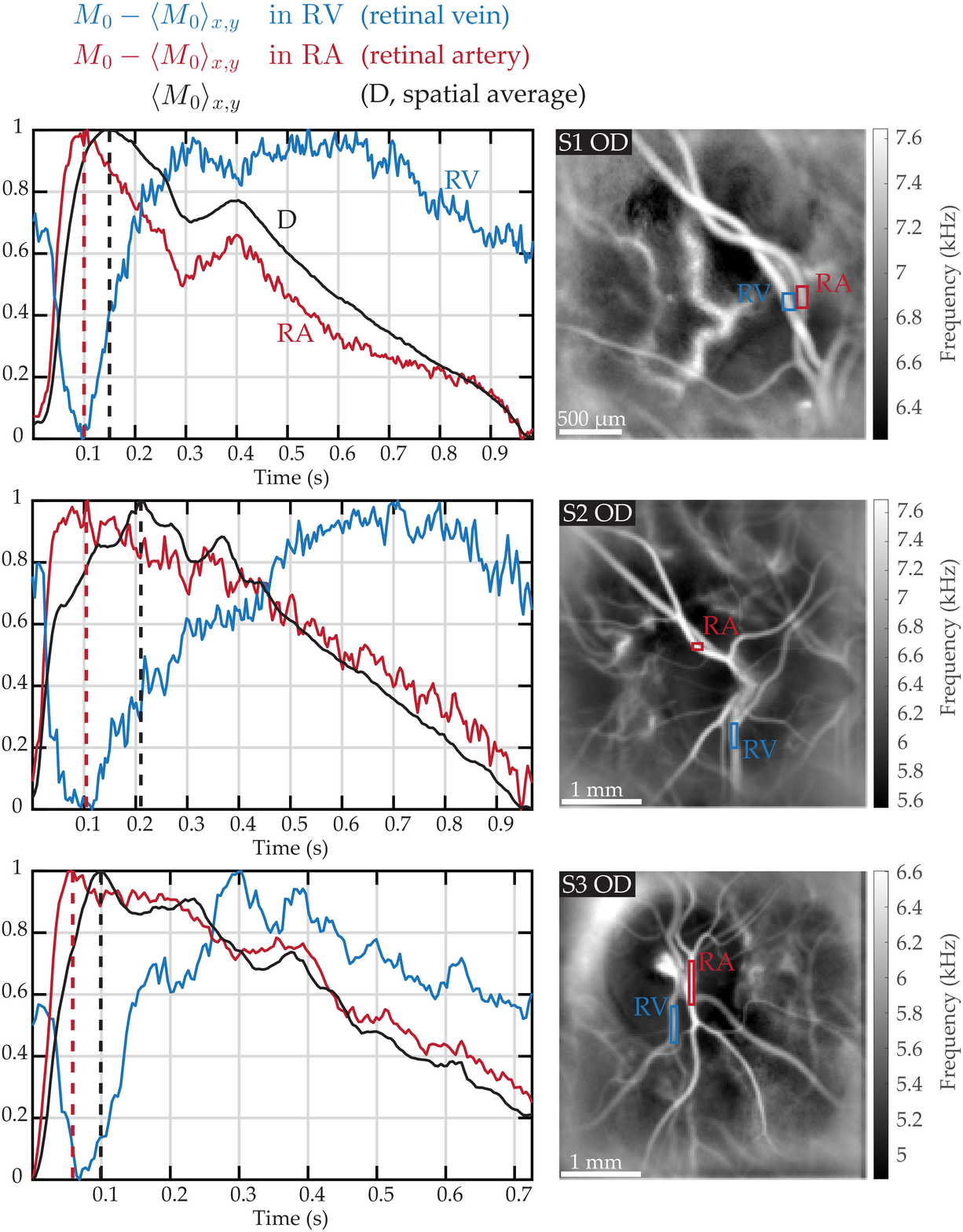}
\caption{Venous and arterial retinal waveforms compared to the spatial average in three examples close to the ONH. Left column: normalized waveform in a retinal vein and artery ('RV' in blue, and 'RA' in red), and normalized waveform of the spatial average (over the entire image) in black. As seen from the red and black dashed lines indicating the end of systole, the venous minimum is more contemporary with the retinal arterial flow maximum than with the maximum of the spatial average. See \textcolor{blue}{\href{https://osapublishing.figshare.com/articles/media/Visualization_4/7988216}{Visualization 3}} and \textcolor{blue}{\href{https://osapublishing.figshare.com/articles/media/Visualization_5/7988315}{Visualization 4}}.
}
\label{fig_8_TriplePlots}
\end{figure}

We applied the waveform normalization process in Fig.~\ref{fig_8_TriplePlots} to investigate how the retinal arterial and venous waveform timings are linked. To this end, we compare waveform profiles normalized between 0 and 1, and corrected from the spatial average in a retinal artery 'RA' (red) and a retinal vein 'RV'(blue), and the normalized waveform of the spatially averaged dominant signal 'D' (black). It should be kept in mind that the amplitude of venous flow variations are in reality considerably lower than the amplitude of arterial flow variations, as was shown in Fig.~\ref{fig_2_HighFrequencyShifts}. The measurements were performed in three different subjects, and averaged over two to four cardiac cycles.

In all the examples shown in Fig.~\ref{fig_8_TriplePlots}, the spatially averaged power Doppler and the power Doppler in the retinal artery follow similar variations, but the systolic peak is earlier by a few tens of milliseconds in the retinal artery with a few inter-individual variations. The arterial and venous waveforms also vary from one person to another but some common features can be observed: right before the systolic peak, a venous drop is synchronous with a rapid arterial increase up to the arterial systolic peak which coincides with the venous minimum flow. After the systole, the arterial flow drops then increases a little at the dicrotic notch (when present), and then slowly decreases until the end of diastole. For its part the venous flow increases after the systole and reaches its maximum a few hundreds milliseconds after the systole, then slowly decreases till the diastole end. When looking more closely, it can be seen that the venous minimum is more contemporary with the arterial maximum than with the dominant signal maximum. In the first example shown in \textcolor{blue}{\href{https://osapublishing.figshare.com/articles/media/Visualization_4/7988216}{Visualization 3}}, it can also be observed that the arterial increase is contemporary with a venous decrease during the dicrotic notch. In \textcolor{blue}{\href{https://osapublishing.figshare.com/articles/media/Visualization_5/7988315}{Visualization 4}}, the lower hemivein swells and retracts in a timing consistent with the power Doppler curve: the vessel retracts until the flow is minimum, and reaches its maximum swelling when the measured power Doppler is maximum.

The dominant signal is subtracted from the dynamic signal, thus it is possible that there is an overcompensation of this signal and which could introduce artifacts in the shape of the venous profile. However it seems that the venous minimum contemporary of the arterial maximum is not an artifact. Indeed in case of an overcompensation, the venous minimum would have been synchronous with the maximum of the dominant signal 'D'. The fact that it is instead synchronous with the arterial signal indicates that it likely truthfully reflects the retinal venous blood flow waveform. Moreover, the caliber of the vein in \textcolor{blue}{\href{https://osapublishing.figshare.com/articles/media/Visualization_5/7988315}{Visualization 4}} changes in a pattern consistent with the flow measurements. This vein's behavior may be particular to that subject due to a specific vascular branching, as one of the vein branches makes an acute angle with the larger vein which might complicate the convergence of venous inflow. Nonetheless, the arterial and venous waveforms measured in these vessels are representative of those we measured in all other eyes.

\subsection{Blood flow resistivity mapping}
\begin{figure}[t!]
\centering
\includegraphics[width = 1\linewidth]{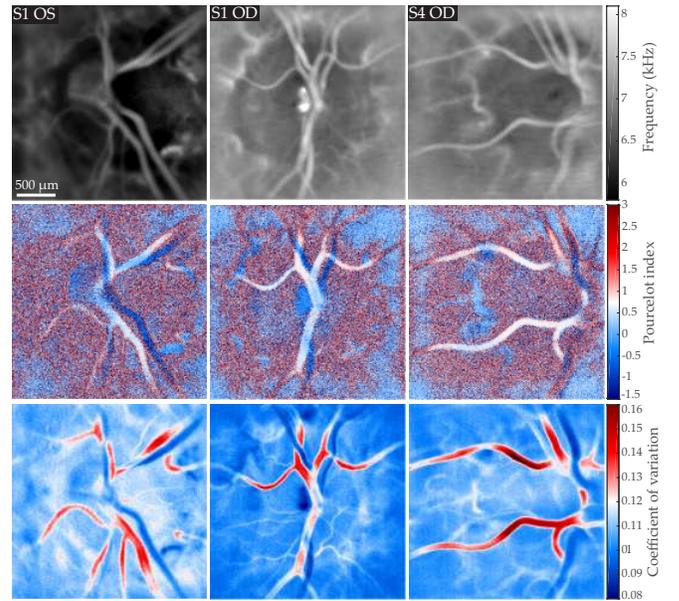}
\caption{Mapping in three different eyes of the local blood velocity, RI, and CV indices; all images are displayed on the same gray/color scale. First row: mean Doppler frequency shift  $f_{\rm mean}(x,y)$. Second row: resistivity map calculated according to the Pourcelot index on the flow variations corrected from the baseline signal $RI_{M_0 - \langle M_{0} \rangle_{x,y}}$; the RI value in the arteries is around 0.7. Third row: coefficient of variation map of the raw signal $CV_{M_0}$.
}\label{fig_3_ResistivityMaps_Retine}
\end{figure}

In Fig.~\ref{fig_3_ResistivityMaps_Retine}, we show in three different eyes examples of mapping of the local blood velocity, RI, and CV indices. For each eye, the measurement was performed in the vicinity of the optic disc, and the camera sampling frequency was set to $f_{S} = 75 \rm kHz$. The first row of Fig.~\ref{fig_3_ResistivityMaps_Retine} shows the local mean Doppler frequency shift $f_{\rm mean}(x,y)$ calculated over the frequency range [0.5-37 kHz]. The second row shows the resistivity map calculated according to the Pourcelot index $RI_{M_0 - \langle M_{0} \rangle_{x,y}}$ from power Doppler movies corrected from the background. Finally, the third row shows the local coefficient of variation $CV_{M_0}$ calculated from the raw power Doppler variations.

On the time-averaged frequency maps $f_{\rm mean}(x,y)$, the mean Doppler frequency shift is greater in greater vessels: the value measured within large retinal vessels is around 7-8 kHz, and 6-6.5 kHz in the optic disc where there are no visible vessels but there should be unresolved capillaries. All $f_{\rm mean}$ images were averaged over $3.5 \rm s$, so it is not possible to distinguish arteries from veins based on the value of the mean frequency shift because when averaged over cardiac cycles these values are too similar in the two types of vessels. With the resistivity maps shown below in Fig.~\ref{fig_3_ResistivityMaps_Retine}, a difference of contrast between veins and arteries appears: with the Pourcelot index, it is possible to see large retinal arteries in white, and large retinal veins in blue. The resisitivity index value measured in these examples is approximately 0.7 for arteries, which is consistent with the values measured with Doppler sonography in the central retinal artery~\cite{Tranquart2003}. The coefficient of variations maps calculated on the raw power Doppler signal show similar contrast but with a much better signal to noise ratio. Arteries come out in red with values around 0.12 - 0.16, veins come out in blue with values in the range of 0.08 - 0.11, and the value in the background is around 0.10.

This difference of contrast between veins and arteries in the RI and CV maps can be easily understood in the light of the information shown in Fig.~\ref{fig_2_HighFrequencyShifts}. Whether the maps are computed using the raw signals or on the signal corrected from the spatial average, the amplitude of systolodiastolic variations is larger in arteries than in veins, thus a difference of contrast appears. The systolodiastolic variations are slightly greater in the veins than in the background, but the variations normalized by the mean value are greater in the background. Consequently the veins come out as the features with the lowest RI and CV values. The most striking difference between $CV_{M_0}$ and $RI_{M_0 - \langle M_{0} \rangle_{x,y}}$ is an improvement of the SNR, which is due to the fact that the coefficient of variation calculation uses all the data points of the cardiac cycles, whereas the Pourcelot index is solely based on two time measurements made at peak systolic and end diastolic time points. Only the Pourcelot index holds a real physical sense, as the values of the coefficient of variations are based on raw power Doppler which are figurative of the background signal variations. However as these two types of maps yield similar contrasts, we used the coefficient of variation in the rest of the manuscript to provide high SNR arteriovenous mapping. Interestingly, with the LDH technique we used in this work, the Doppler contrast is not quite equally sensitive to in-plane motion as it is to out of plane motion~\cite{Puyo2018}. This may result in overestimation of blood volume and velocity in out of plane vessels. The resistivity maps are predicted to be relatively independent of the flow geometry given that the angular Doppler sensitivity is expected to be compensated in the ratio. Finally it can be noticed that resistivity values in arteries are corrupted when there is a vein beneath. This is due to the relative transparency of blood vessels and the large depth of field of the instrument; it can be also be noticed on flow images that when vessels overlap their intensity is cumulated.

\subsection{Comparative CV mapping with $f_{\rm mean}$ and $M_{\rm 0}$}
\begin{figure}[t!]
\centering
\includegraphics[width = 1\linewidth]{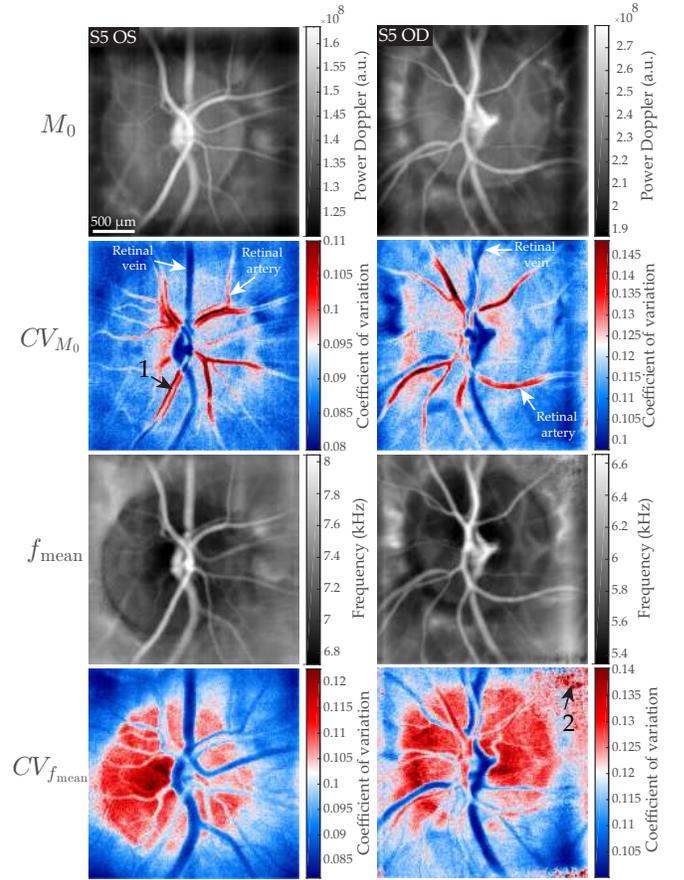}
\caption{Comparison of CV maps obtained from the variation of $M_{0}$ and $f_{\rm mean}$ in the right and left eye of the same subject. From top to down: $M_{0}(x,y)$, $CV_{M_{0}}(x,y)$, $f_{\rm mean}(x,y)$, and $CV_{f_{\rm mean}}(x,y)$. The coefficient of variation in the optic disc differs significantly between the two methods. The arrow '1' and '2' point to areas where the Doppler broadening is undersampled and to low frequency noise, respectively.
}\label{fig_3_ResistivityMaps_Retine_M1sM0}
\end{figure}

In Fig.~\ref{fig_3_ResistivityMaps_Retine_M1sM0}, we compare the coefficient of variation maps that can be obtained on the basis of the pixel-wise variations of $f_{\rm mean}$, and compare them to the maps calculated from the variations of $M_{\rm 0}$ as those that were shown in Fig.~\ref{fig_3_ResistivityMaps_Retine}. The two corresponding maps are denoted $CV_{f_{\rm mean}}(x,y)$ and $CV_{M_{0}}(x,y)$. The two columns in Fig.~\ref{fig_3_ResistivityMaps_Retine_M1sM0} show these images for the right and left eye of the same subject; both measurements were performed with a camera frame rate of 60 kHz. The frequency ranges used for the calculations of $f_{\rm mean}$ and $M_{\rm 0}$ are 0.5-30 kHz and 6-30 kHz, respectively.

With both the $CV_{f_{\rm mean}}$ and $CV_{M_{0}}$ maps, a difference of contrast is visible between arteries and veins, and the values found in $CV_{M_{0}}(x,y)$ are very close to those of $CV_{f_{\rm mean}}(x,y)$. The most visible difference is that in $CV_{f_{\rm mean}}(x,y)$ maps, the coefficient of variation is greater in the optic disc than in almost any other feature, except for arteries in the optic disc because they benefit from the signal of the optic disc underneath. As a result, unlike with $CV_{M_{0}}(x,y)$ map, retinal arteries are not the structures where the coefficient of variation is the greatest. Compared to the rest of the visible features, the optic disc area appears darker with $f_{\rm mean}(x,y)$ than with $M_{0}(x,y)$. As $CV_{f_{\rm mean}}(x,y)$ is calculated as the division of the standard deviation image by the time-averaged image, this explains why this region appears redder than the rest of the image in the coefficient of variation maps. As for the retinal veins, they are the structures with the lowest coefficient of variation with the two types of maps.
It can also be noticed that the apparent diameter of retinal vessels is larger with images based on $f_{\rm mean}$ than with images based on $M_{\rm 0}$. A likely explanation is that because the 0.5-6 kHz frequency range is not used in $M_{\rm 0}$ images, the contribution of higher frequencies (which reveal greater blood flow) is more important, so vessels show a weaker signal closer to the edges.

It can be noted than the lumen of some large retinal arteries in $CV_{M_{0}}(x,y)$ (arrow '1') has a lower coefficient of variation than closer to the vessel walls. As we fail to see any physical explanation of this phenomenon, we suppose it is caused by an aliasing of the Doppler spectrum during systole in the area where the velocity is the greatest, which would result in a lower than normal systolic power Doppler flow measurement, and thus in lower systolodiastolic variations. We had already noticed that at 60 kHz this sort of Doppler undersampling could occur in branches of the central retinal artery~\cite{Puyo2018}. An interesting feature of $CV_{f_{\rm mean}}(x,y)$ maps, is that as $f_{\rm mean}$ itself, they are more prone to low frequency noise (arrow '2'), and require a careful instrumentation that minimizes specular reflections.

\section{Results in the choroid} \label{section_choroid}
The blood flow supply to the retina is carried out by two independent vasculatures originating from the ophthalmic artery that differ in organization and function. The retinal vessels are branches of the central retinal artery and supply the neural retina and the prelaminar region of the ONH, whereas the choroidal vessels originate from the posterior ciliary arteries (PCAs) which supply the photoreceptor and epithelial cells. These two vascular systems especially differ at the capillary bed level as choroidal capillaries are considerably bigger than retinal capillaries~\cite{Anand2010}, resulting in a lower vascular resistance in the choroid. Therefore, because they share a common inflow source but differ in outflow, the retinal and choroidal circulation are interesting to study the ocular circulation. The difficulty about the choroid is that it lies beneath the photoreceptor and RPE layers which have a non-trivial effect on the light. We previously demonstrated that LDH is able to reveal the deep choroidal vasculature, and that this difference of organization allows to discriminate arteries from veins in the choroid on the basis of the flow they carry~\cite{Puyo2019}. Here we investigate the blood flow dynamics in the choroid.

\subsection{Choroidal arterial waveform in different regions}
\begin{figure}[t!]
\centering
\includegraphics[width = 1\linewidth]{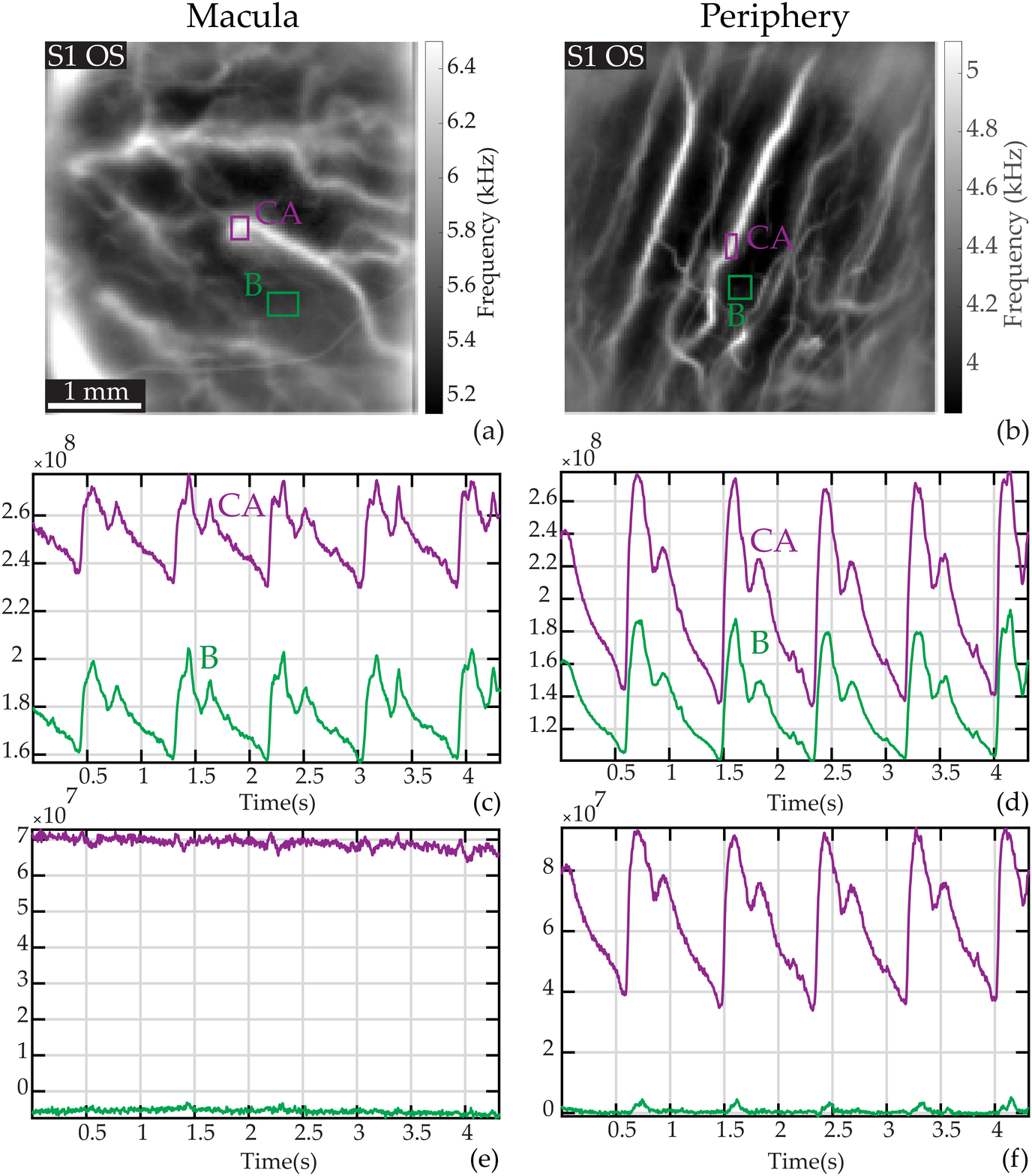}
\caption{Laser Doppler measurements in choroidal arteries in the macular and peripheral regions. (a) and (b): Mean frequency images with ROIs on a large choroidal artery ('CA', purple) and the background ('B', green). (c)  and (d): Raw power Doppler signal $M_{0}$ in the ROIs. (e) and (f): Power Doppler signal in the same ROIs when subtracting the spatially averaged dynamic value, i.e., $M_{0}- \langle M_{0} \rangle_{x,y}$. In the peripheral region the power Doppler signal in 'CA' corrected from the dominant signal yields a arterial-like waveform whereas the power Doppler signal measured in 'CA' in the macular region shows only noise. This is probably because of the higher photoreceptors/RPE density, and because the dynamic signal measured in 'CA' in the macula is exactly like the spatial average signal.
}
\label{fig_4_zChoroidPlot}
\end{figure}

In Fig.~\ref{fig_4_zChoroidPlot}, we explore how the photoreceptor density affects the laser Doppler measurements in choroidal arteries. To that end, we imaged two areas in the fundus, in the macular region and in a peripheral region at approximately 40 degrees of foveal eccentricity where the density of photoceptor and RPE cells is expected to be considerably lower. In Fig.~\ref{fig_4_zChoroidPlot}(a) and (b), we show the mean frequency images of the two imaged areas, and the ROIs placed on a choroidal artery ('CA', purple) and an area devoid of any large vessels ('B', green). In Fig.~\ref{fig_4_zChoroidPlot}(c) and (d), we show the raw power Doppler measured in the ROIs, i.e. $M_{0}$. Then in Fig.~\ref{fig_4_zChoroidPlot}(e) and (f), we show the power Doppler in the same ROIs when subtracting the spatially averaged dynamic signal, i.e. $M_{0} - \langle M_{0} \rangle_{x,y}$.

Similarly to the results shown in Fig.~\ref{fig_2_HighFrequencyShifts}, the raw power Doppler variations in the two ROIs have very correlated variations. When subtracting the baseline dynamic value, a pulsatile arterial signal remains in the choroidal artery in the peripheral region, whereas in the macular region the signal measured in the choroidal artery only shows noise. We presume it is the higher density of photoreceptor and RPE cells in the macular region that prevents us from making measurements in choroidal arteries in the same way as in the periphery. A plausible explanation is that the choroidal arterial signal in the macular region is exactly the dominant signal, which explains why it is totally suppressed when subtracting the spatially averaged signal. This particularity had already been demonstrated in laser Doppler flowmetry as the absence of retinal vasculature in the fovea was already used to make choroidal blood flow measurements~\cite{Riva1994}.

\subsection{Simultaneous blood flow measurements in the retina and choroid}
\begin{figure}[t!]
\centering
\includegraphics[width = 1\linewidth]{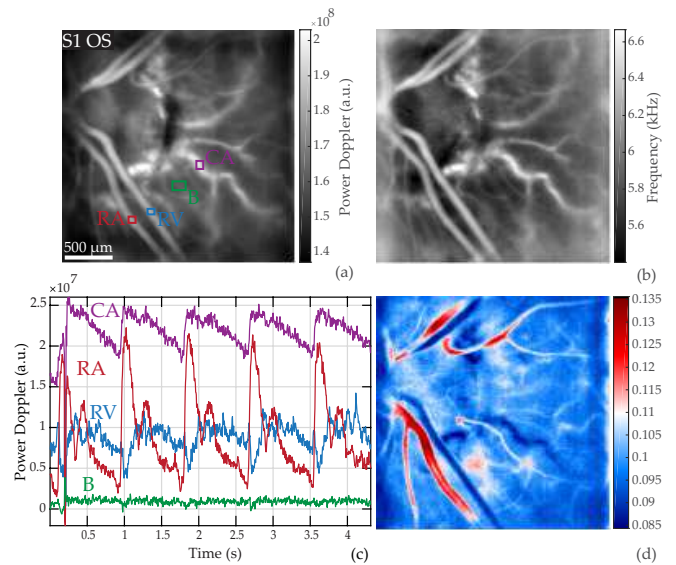}
\caption{Pulsatile flow in the retina and choroid. (a) Power Doppler image $M_{0}(x,y)$ indicating the ROIs; the red, blue, green, purple boxes mark a retinal artery and vein, the background, and a choroidal artery, respectively. (b) Mean frequency shift image $f_{\rm mean}(x,y)$. (c) Waveform of $M_{0}- \langle M_{0} \rangle_{x,y}$ in the ROIs. (d) Coefficient of variation map $CV_{M_{0}}(x,y)$: the choroidal artery has a coefficient of variation close to the values found in retinal veins, despite exhibiting an arterial waveform profile. See the power Doppler movie in \textcolor{blue}{\href{https://osapublishing.figshare.com/articles/media/Visualization_3/7988222}{Visualization 5}}.
}
\label{fig_4_ResistivityMaps_Choroid_Plot}
\end{figure}

In Fig.~\ref{fig_4_ResistivityMaps_Choroid_Plot}, we provide an example of a CV map in an area close to the ONH with visible choroidal arteries that are branches of paraoptic short PCAs. The camera sampling frequency was set to 60 kHz. In Fig.~\ref{fig_4_ResistivityMaps_Choroid_Plot}(a) and (b) are respectively shown the time-averaged power Doppler image $M_{0}(x,y)$ and mean Doppler frequency shift image $f_{\rm mean}(x,y)$. The power Doppler waveforms in the ROIs corrected from the spatial average are plotted in Fig.~\ref{fig_4_ResistivityMaps_Choroid_Plot}(c), and finally Fig.~\ref{fig_4_ResistivityMaps_Choroid_Plot}(d) shows the coefficient of variation map $CV_{M_{0}}(x,y)$. The corresponding power Doppler movie corrected from the spatial average is shown in \textcolor{blue}{\href{https://osapublishing.figshare.com/articles/media/Visualization_3/7988222}{Visualization 5}}.

We see here the same distinct flow behaviors in the retinal artery and vein as before, leading to similar contrast on the CV map. The choroid artery 'CA' appears on the power Doppler and mean frequency images with a Doppler signal greater than in any other feature. The blood flow measured in this vessel in Fig.~\ref{fig_4_ResistivityMaps_Choroid_Plot}(c) exhibits an arterial-like waveform with very reduced systolodiastolic variations compared to the retinal artery, and slightly more noisy; it seems to have an offset value compared to the retinal vessel. it is apparent in \textcolor{blue}{\href{https://osapublishing.figshare.com/articles/media/Visualization_3/7988222}{Visualization 5}} that the measured pulsatility in the choroidal artery is very reduced compared to that of the retinal artery. Finally in the CV map displayed in Fig.~\ref{fig_4_ResistivityMaps_Choroid_Plot}(d), the choroidal artery 'CA' can be seen with a similar contrast to that of retinal veins.

This lower modulation depth combined with the high average value explains why this choroidal artery appears in blue in the coefficient of variation map. Although reduced systolodiastolic variations are expected in the choroid because of the lower vascular resistance compared to the retina, the photoreceptor and RPE could also be playing a role in reducing the amplitude due the multiple scattering and absorption they introduce. This makes it impossible to compare the raw power Doppler value between the retinal and choroidal layers. To circumvent this issue, we here limit the study to the comparisons of shapes and timings of choroidal and retinal arteries flow waveforms in Fig.~\ref{fig_7_TriplePlots_RetinaChoroid} by normalizing the waveform profiles.

\subsection{Comparison of the waveform of retinal arteries, choroidal arteries, and of the spatial average}
\begin{figure}[t!]
\centering
\includegraphics[width = 1\linewidth]{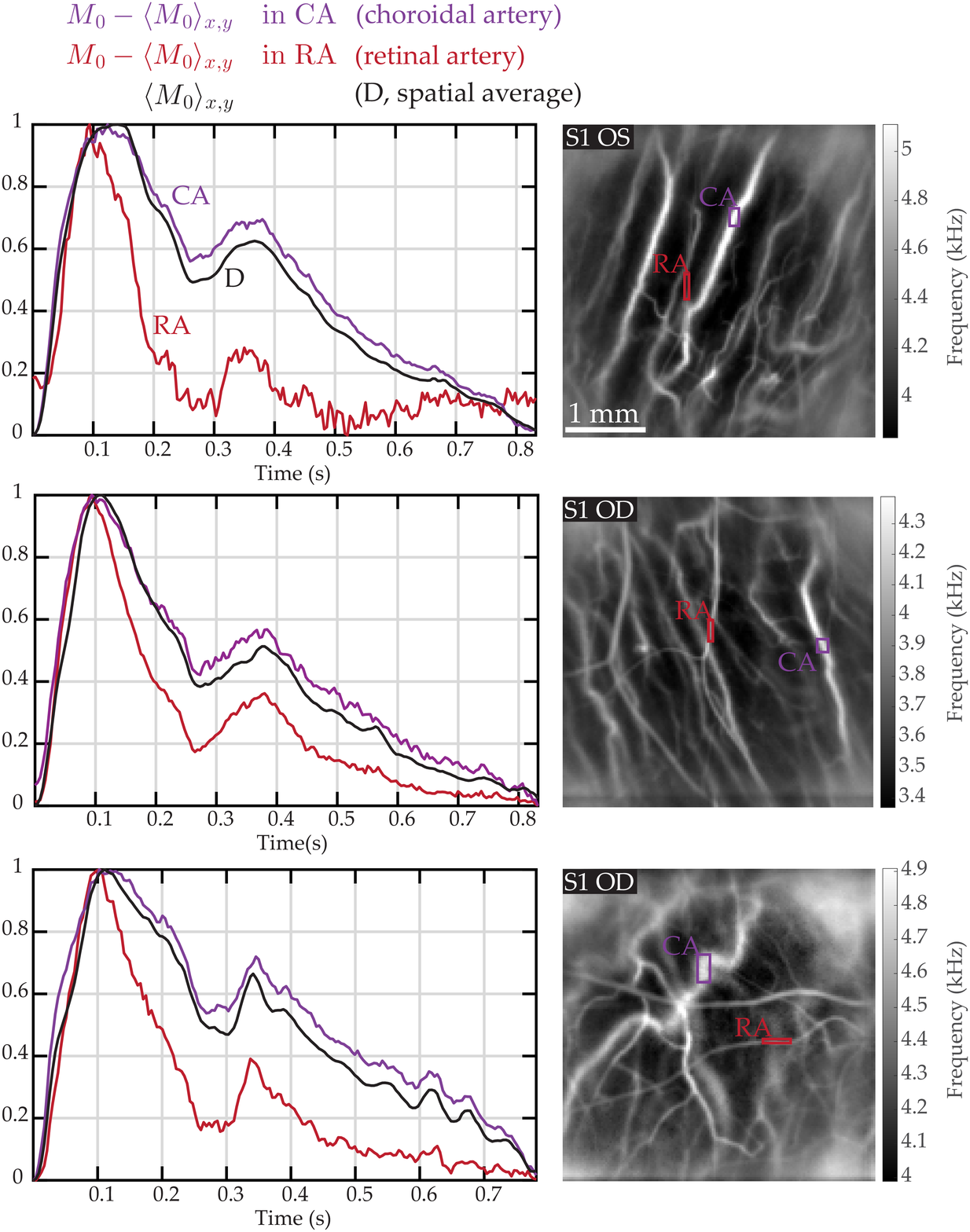}
\caption{Retinal and choroidal arterial waveforms compared to the spatial average in three examples in peripheral regions. Left column: normalized waveform of power Doppler corrected from the spatial average in a retinal and a choroidal artery ('RA' in red, and 'CA' in purple), and normalized waveform of the Doppler signal spatially averaged over the entire image ('D' in black). The dominant signal waveform has an arterial waveform, which strongly resembles that of choroidal arteries.
}
\label{fig_7_TriplePlots_RetinaChoroid}
\end{figure}

In Fig.~\ref{fig_7_TriplePlots_RetinaChoroid}, we compare the waveform profiles of retinal and choroidal arteries, and of the dominant signal. In three examples, we measured the power Doppler signal corrected from the spatial average in a retinal and a choroidal artery ('RA' and 'CA'), and averaged the waveform profiles over several cardiac cycles (typically 2 to 4). We then compared these waveforms to the spatial average waveform 'D', i.e. to $\langle M_{0} \rangle_{x,y}$. All waveforms were normalized to have variations between 0 and 1. For each example, we show the ROIs on the mean frequency shift image on the right, and the corresponding curves on the left. The retinal artery is marked 'RA' in red, and the choroidal artery 'CA' in purple.

In the three examples, all waveforms exhibit an arterial profile: the systolic, diastolic, and dicrotic times are visible. No difference of synchronicity for the start of the systole and dicrotic notch can be observed between the different waveforms, but the systolic peak is slightly delayed in 'D' and 'CA'. A dominant contribution to the spatial average of light scattered in the choroid seems likely.

\subsection{Arteriovenous differentiation in the retina and choroid}
\begin{figure}[t!]
\centering
\includegraphics[width = 1\linewidth]{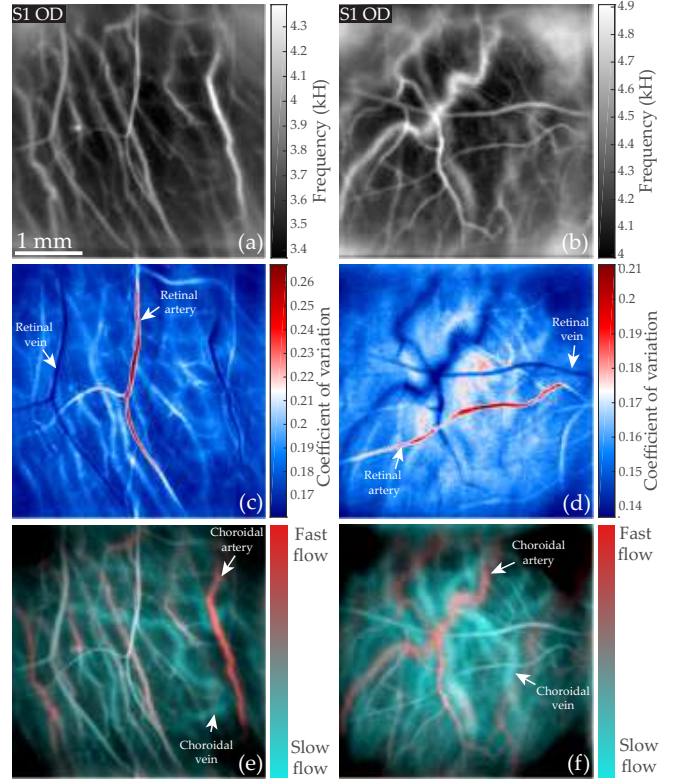}
\caption{Arteriovenous differentiation in the retina and choroid. (a) and (b): mean Doppler frequency $f_{\rm mean}(x,y)$. (c) and (d): coefficient of variation $CV_{M_0}$ revealing retinal arteries and veins. (e) and (f): color composite images obtained by merging low frequency and high (2.5-6 kHz, 10-30 kHz) power Doppler images in cyan and red, allowing identification of choroidal arteries and veins (colorbars are only indicative). 
}\label{fig_5_ResistivityMaps_Choroid_Peripherie}
\end{figure}

In Fig.~\ref{fig_5_ResistivityMaps_Choroid_Peripherie}, we demonstrate the ability of LDH to perform an arteriovenous differentiation in the retina and choroid using a method adapted to each vasculature. In Fig.~\ref{fig_5_ResistivityMaps_Choroid_Peripherie}(a) and (b) we show images of the local mean Doppler frequency shift $f_{\rm mean}(x,y)$. Then in Fig.~\ref{fig_5_ResistivityMaps_Choroid_Peripherie}(c) and (d), we show the mapping of the local coefficient of variation $CV_{\rm M_{0}}(x,y)$. Finally, in Fig.~\ref{fig_5_ResistivityMaps_Choroid_Peripherie}(e) and (f), we show the composite color images obtained by merging low and high frequency power Doppler images (calculated with the frequency ranges 2.5-6, and 10-30 kHz, respectively) in cyan and red, following the process we previously demonstrated~\cite{Puyo2019}.

In the mean frequency shift images, all vessels from both the retinal and choroidal layers are visible except the choroidal veins. Although one can distinguish the retinal from the choroidal vessels as the latter are slightly larger, identifying the type of each vessel on the basis on the mean frequency shift is not possible. In the coefficient of variation maps, a difference of contrast appears between retinal arteries and veins and they can be clearly identified: as we have shown before, arteries come out in bright red and veins in deep blue. However on $CV$ maps, the vessels of the choroid cannot be clearly identified: choroidal arteries have the same contrast as retinal veins, and choroidal veins have the same contrast as the background. Finally in the composite color images, while the retinal vessels appear with a neutral contrast, large differences can be noticed among choroidal vessels: choroidal arteries clearly stand out in red because they carry a larger flow, while choroidal veins can be seen in cyan. A final observation regarding Fig.~\ref{fig_5_ResistivityMaps_Choroid_Peripherie}, is that in these peripheral regions, the mean frequency shift is lower here than in regions close to the ONH, while the average coefficient of variation index is greater.

LDH is able to carry out an arteriovenous differentiation of all vessels because it is able to make quantitative measurements of power Doppler in both the retina and choroid, and this quantity is related to the flow. For the retinal vasculature the discrimination is made on the basis of the respective hemodynamics of arteries and veins using CV maps. In the choroid, the arteriovenous segregation relies on the ability to analyze the flow in choroidal vessels. Choroidal venous vessels are the hardest vessels to image in laser Doppler because they carry a smaller flow. We had found that in regions of eccentricity smaller than approximately 30-40 degrees, the Doppler broadening due to blood flow in choroidal veins lies in the frequency range 2.5-6 kHz~\cite{Puyo2019}. However in this same range of frequency there are strong contributions of involuntary eye movements and fundus pulsations, which prevents us from measuring a dynamic signal. This is why so far we have not managed to measure the blood flow waveform in choroidal veins.

\section{Discussion and conclusion} \label{sec_Discussion}

Our results indicate that the choroid constitutes the predominant contribution to the high frequency laser Doppler signal in all reported areas of the fundus. It is however possible to circumvent its influence by subtracting the spatially averaged baseline signal. Similar conclusions have been reached with other blood flow monitoring methods. In laser speckle flowgraphy, the spatially averaged value of MBR (mean blur rate) which is used to assess blood flow is subtracted in order to correct for the background pulsatile signal due to the choroid~\cite{Isono2003, Iwase2015, Fondi2018}. In scanning LDF, it was found that the measured perfusion maps did not give an accurate description of the retinal circulation alone as they failed to predict retinal hypoperfusion where expected~\cite{Squirrell2001}. The conclusion reached was that despite confocal gating, laser Doppler measurements represented the combined circulations of the retina and choriocapillaris. The ability to make full-field laser Doppler measurements thus appears as crucial as it allows to filter this dominant contribution from the choroid through the subtraction of the spatial average. This overall signal is most likely due to the scattering of the light backscattered by the choroid by the intermediate layers, with a likely Doppler broadening contribution from the choroidal precapillary sphincter or choriocapillaris that would explain why the spatially averaged signal reaches its systolic peak later than the retinal arteries.

In Fig.~\ref{fig_7_TriplePlots_RetinaChoroid}, we observed different waveform profiles in retinal and choroidal vessels. These measurements could potentially reflect a difference of arterial waveforms between these two vasculatures, but the important bias introduced by the high-pass frequency filtering of the power Doppler prevents any direct conclusion. As we saw in Fig.~\ref{fig_2_SeveralFrequencyRanges}, the spectral contribution of global eye movements overlaps with the Doppler broadening due to pulsatile blood flow in the 6-10 kHz frequency range, especially in the periphery where retinal vessels are smaller and carry slower flows. We performed these measurements shown in Fig.~\ref{fig_7_TriplePlots_RetinaChoroid} in the periphery to make measurement in choroidal vessels with a higher SNR, but the retinal arteries are smaller thus strongly impacted by the high-pass Doppler frequency filter set at 10 kHz. The frequency thresholding acts as a high-pass velocity filter which leads to a sharper waveform profile as the contribution of the constant flow is removed. Thus the pulsatile shape of the retinal arteries waveform profiles in Fig.~\ref{fig_7_TriplePlots_RetinaChoroid} is amplified. Another doubt can be retained regarding the exactitude of the measured choroidal arterial waveform as it is possible that there remains a residual contribution of the dominant signal. Thus any interpretation based on the comparison of retinal and choroidal arterial waveforms should be made cautiously. Nonetheless these results stand in good agreement with the profiles measured with Doppler sonography measurements~\cite{Tranquart2003, Lieb1991}, as it was found that the flow in PCAs exhibits lower systolodiastolic variations than in the central retinal artery, as it is typical in case of a lower downstream vascular resistance. LDH may be a useful technique to further study how the distal vascular resistance alters the arterial waveform.

We observed pulsatile blood flow changes in retinal veins in all investigated eyes. \textcolor{blue}{\href{https://osapublishing.figshare.com/articles/media/Visualization_5/7988315}{Visualization 4}} shows spontaneous retinal venous pulsation (SRVP). This is commonly observed in normal veins close to the disc, and probably results from interactions between intraocular and intracranial pressure (ICP), although the exact mechanism is uncertain~\cite{Morgan2016, Wartak2019}. Our data suggests that the peak venous diameter is associated with the maximum of flow. Given the interest of such phenomenon for probing ICP, further research are warranted to better understand the flow behavior in cases of SRVP. Venous pulsation is rather uncommon in the human body, and there has been controversial reports regarding the phase of the flow and caliber pulsatile behavior of retinal veins~\cite{Kain2010, Morgan2012, Morgan2014, Moret2015}. In this work we have found the arterial blood flow systolic peak to be coincident with the venous minimum flow in all investigated eyes. In the past decades, several modern ophthalmic instruments have experimentally confirmed the coincidence of arterial maximum flow and venous minimum flow such as Doppler-OCT~\cite{Tan2015, Doblhoff2014}, Doppler sonography~\cite{Michelson1997, Tranquart2003}, or dynamic angiography in cases of retinal vein occlusion~\cite{Paques2005}.

In conclusion, we have shown that thanks to its high temporal resolution and full-field imaging capability, LDH is able to measure blood flow in retinal arteries and veins, and exhibit their specifically interrelated waveforms. Further we show that despite the absence of sectioning ability in LDH which results in having the retinal and choroidal layers reconstructed in a single plane, by spatio-temporal filtering it seems possible to isolate the contributions of each vasculature. We also demonstrate the use of the resistivity index which leads to a clear arteriovenous differentiation in the retina that complements the ability of LDH to differentiate arteries from veins in the choroid that we had previously demonstrated. The ability to characterize retinal and choroidal hemodynamic in the normal physiological state with LDH lays the foundation for studying altered blood flow in pathological conditions.

\section*{Funding}
This work was supported by LABEX WIFI (Laboratory of Excellence ANR-10-LABX-24) within the French Program Investments for the Future under Reference ANR-10-IDEX-0001-02 PSL, and the European Research Council (ERC Synergy HELMHOLTZ, grant agreement \#610110). The Titan Xp used for this research was donated by the NVIDIA Corporation.

\section*{Acknowledgments}
The authors would like to thank Sarah Mrejen, and Philippe Bonnin for helpful discussions, Kate Grieve for language corrections and help in obtaining safety authorizations, and reviewers for their helpful comments.

\section*{Disclosures}
The authors declare that there are no conflicts of interest related to this article.

\section*{Supplementary Material}
\noindent
\textcolor{blue}{\href{https://youtu.be/BairRa4aJxU}{Supplementary Visualization 1}}. \newline

\bibliography{./Bibliography}

\begin{thebibliography}{66}
\expandafter\ifx\csname natexlab\endcsname\relax\def\natexlab#1{#1}\fi
\expandafter\ifx\csname bibnamefont\endcsname\relax
  \def\bibnamefont#1{#1}\fi
\expandafter\ifx\csname bibfnamefont\endcsname\relax
  \def\bibfnamefont#1{#1}\fi
\expandafter\ifx\csname citenamefont\endcsname\relax
  \def\citenamefont#1{#1}\fi
\expandafter\ifx\csname url\endcsname\relax
  \def\url#1{\texttt{#1}}\fi
\expandafter\ifx\csname urlprefix\endcsname\relax\def\urlprefix{URL }\fi
\providecommand{\bibinfo}[2]{#2}
\providecommand{\eprint}[2][]{\url{#2}}

\bibitem[{\citenamefont{Kashani et~al.}(2017)\citenamefont{Kashani, Chen, Gahm,
  Zheng, Richter, Rosenfeld, Shi, and Wang}}]{Kashani2017}
\bibinfo{author}{\bibfnamefont{A.~H.} \bibnamefont{Kashani}},
  \bibinfo{author}{\bibfnamefont{C.-L.} \bibnamefont{Chen}},
  \bibinfo{author}{\bibfnamefont{J.~K.} \bibnamefont{Gahm}},
  \bibinfo{author}{\bibfnamefont{F.}~\bibnamefont{Zheng}},
  \bibinfo{author}{\bibfnamefont{G.~M.} \bibnamefont{Richter}},
  \bibinfo{author}{\bibfnamefont{P.~J.} \bibnamefont{Rosenfeld}},
  \bibinfo{author}{\bibfnamefont{Y.}~\bibnamefont{Shi}}, \bibnamefont{and}
  \bibinfo{author}{\bibfnamefont{R.~K.} \bibnamefont{Wang}},
  \bibinfo{journal}{Progress in Retinal and Eye Research}
  (\bibinfo{year}{2017}).

\bibitem[{\citenamefont{Chua et~al.}(2019)\citenamefont{Chua, Chin, Hong, Chee,
  Le, Ting, Wong, and Schmetterer}}]{ChuaChin2019}
\bibinfo{author}{\bibfnamefont{J.}~\bibnamefont{Chua}},
  \bibinfo{author}{\bibfnamefont{C.~W.~L.} \bibnamefont{Chin}},
  \bibinfo{author}{\bibfnamefont{J.}~\bibnamefont{Hong}},
  \bibinfo{author}{\bibfnamefont{M.~L.} \bibnamefont{Chee}},
  \bibinfo{author}{\bibfnamefont{T.-T.} \bibnamefont{Le}},
  \bibinfo{author}{\bibfnamefont{D.~S.~W.} \bibnamefont{Ting}},
  \bibinfo{author}{\bibfnamefont{T.~Y.} \bibnamefont{Wong}}, \bibnamefont{and}
  \bibinfo{author}{\bibfnamefont{L.}~\bibnamefont{Schmetterer}},
  \bibinfo{journal}{Journal of Hypertension} \textbf{\bibinfo{volume}{37}},
  \bibinfo{pages}{572} (\bibinfo{year}{2019}).

\bibitem[{\citenamefont{Mrejen and Spaide}(2013)}]{Mrejen2013}
\bibinfo{author}{\bibfnamefont{S.}~\bibnamefont{Mrejen}} \bibnamefont{and}
  \bibinfo{author}{\bibfnamefont{R.~F.} \bibnamefont{Spaide}},
  \bibinfo{journal}{Survey of Ophthalmology} \textbf{\bibinfo{volume}{58}},
  \bibinfo{pages}{387} (\bibinfo{year}{2013}).

\bibitem[{\citenamefont{Rosenfeld et~al.}(2016)\citenamefont{Rosenfeld, Durbin,
  Roisman, Zheng, Miller, Robbins, Schaal, and Gregori}}]{Rosenfeld2016}
\bibinfo{author}{\bibfnamefont{P.~J.} \bibnamefont{Rosenfeld}},
  \bibinfo{author}{\bibfnamefont{M.~K.} \bibnamefont{Durbin}},
  \bibinfo{author}{\bibfnamefont{L.}~\bibnamefont{Roisman}},
  \bibinfo{author}{\bibfnamefont{F.}~\bibnamefont{Zheng}},
  \bibinfo{author}{\bibfnamefont{A.}~\bibnamefont{Miller}},
  \bibinfo{author}{\bibfnamefont{G.}~\bibnamefont{Robbins}},
  \bibinfo{author}{\bibfnamefont{K.~B.} \bibnamefont{Schaal}},
  \bibnamefont{and} \bibinfo{author}{\bibfnamefont{G.}~\bibnamefont{Gregori}},
  in \emph{\bibinfo{booktitle}{OCT Angiography in Retinal and Macular
  Diseases}} (\bibinfo{publisher}{Karger Publishers}, \bibinfo{year}{2016}),
  vol.~\bibinfo{volume}{56}, pp. \bibinfo{pages}{18--29}.

\bibitem[{\citenamefont{Shiga et~al.}(2016)\citenamefont{Shiga, Kunikata,
  Aizawa, Kiyota, Maiya, Yokoyama, Omodaka, Takahashi, Yasui, Kato
  et~al.}}]{Shiga2016}
\bibinfo{author}{\bibfnamefont{Y.}~\bibnamefont{Shiga}},
  \bibinfo{author}{\bibfnamefont{H.}~\bibnamefont{Kunikata}},
  \bibinfo{author}{\bibfnamefont{N.}~\bibnamefont{Aizawa}},
  \bibinfo{author}{\bibfnamefont{N.}~\bibnamefont{Kiyota}},
  \bibinfo{author}{\bibfnamefont{Y.}~\bibnamefont{Maiya}},
  \bibinfo{author}{\bibfnamefont{Y.}~\bibnamefont{Yokoyama}},
  \bibinfo{author}{\bibfnamefont{K.}~\bibnamefont{Omodaka}},
  \bibinfo{author}{\bibfnamefont{H.}~\bibnamefont{Takahashi}},
  \bibinfo{author}{\bibfnamefont{T.}~\bibnamefont{Yasui}},
  \bibinfo{author}{\bibfnamefont{K.}~\bibnamefont{Kato}}, \bibnamefont{et~al.},
  \bibinfo{journal}{Current Eye Research} \textbf{\bibinfo{volume}{41}},
  \bibinfo{pages}{1447} (\bibinfo{year}{2016}).

\bibitem[{\citenamefont{Mursch-Edlmayr
  et~al.}(2018)\citenamefont{Mursch-Edlmayr, Luft, Podkowinski, Ring,
  Schmetterer, and Bolz}}]{Mursch2018}
\bibinfo{author}{\bibfnamefont{A.~S.} \bibnamefont{Mursch-Edlmayr}},
  \bibinfo{author}{\bibfnamefont{N.}~\bibnamefont{Luft}},
  \bibinfo{author}{\bibfnamefont{D.}~\bibnamefont{Podkowinski}},
  \bibinfo{author}{\bibfnamefont{M.}~\bibnamefont{Ring}},
  \bibinfo{author}{\bibfnamefont{L.}~\bibnamefont{Schmetterer}},
  \bibnamefont{and} \bibinfo{author}{\bibfnamefont{M.}~\bibnamefont{Bolz}},
  \bibinfo{journal}{Scientific Reports} \textbf{\bibinfo{volume}{8}},
  \bibinfo{pages}{5343} (\bibinfo{year}{2018}).

\bibitem[{\citenamefont{Bonnin et~al.}(2014)\citenamefont{Bonnin, Pournaras,
  Makowiecka, Krivosic, Kedra, Le~Gargasson, Gaudric, Levy, Cohen, Tadayoni
  et~al.}}]{Bonnin2014}
\bibinfo{author}{\bibfnamefont{P.}~\bibnamefont{Bonnin}},
  \bibinfo{author}{\bibfnamefont{J.-A.~C.} \bibnamefont{Pournaras}},
  \bibinfo{author}{\bibfnamefont{K.}~\bibnamefont{Makowiecka}},
  \bibinfo{author}{\bibfnamefont{V.}~\bibnamefont{Krivosic}},
  \bibinfo{author}{\bibfnamefont{A.~W.} \bibnamefont{Kedra}},
  \bibinfo{author}{\bibfnamefont{J.-F.} \bibnamefont{Le~Gargasson}},
  \bibinfo{author}{\bibfnamefont{A.}~\bibnamefont{Gaudric}},
  \bibinfo{author}{\bibfnamefont{B.~I.} \bibnamefont{Levy}},
  \bibinfo{author}{\bibfnamefont{Y.~S.} \bibnamefont{Cohen}},
  \bibinfo{author}{\bibfnamefont{R.}~\bibnamefont{Tadayoni}},
  \bibnamefont{et~al.}, \bibinfo{journal}{Acta Ophthalmologica}
  \textbf{\bibinfo{volume}{92}}, \bibinfo{pages}{e382} (\bibinfo{year}{2014}).

\bibitem[{\citenamefont{McVeigh et~al.}(2007)\citenamefont{McVeigh, Bank, and
  Cohn}}]{McVeigh2007}
\bibinfo{author}{\bibfnamefont{G.~E.} \bibnamefont{McVeigh}},
  \bibinfo{author}{\bibfnamefont{A.~J.} \bibnamefont{Bank}}, \bibnamefont{and}
  \bibinfo{author}{\bibfnamefont{J.~N.} \bibnamefont{Cohn}}, in
  \emph{\bibinfo{booktitle}{Cardiovascular Medicine}}
  (\bibinfo{publisher}{Springer}, \bibinfo{year}{2007}), pp.
  \bibinfo{pages}{1811--1831}.

\bibitem[{\citenamefont{Avolio et~al.}(2009)\citenamefont{Avolio, Butlin, and
  Walsh}}]{Avolio2009}
\bibinfo{author}{\bibfnamefont{A.~P.} \bibnamefont{Avolio}},
  \bibinfo{author}{\bibfnamefont{M.}~\bibnamefont{Butlin}}, \bibnamefont{and}
  \bibinfo{author}{\bibfnamefont{A.}~\bibnamefont{Walsh}},
  \bibinfo{journal}{Physiological Measurement} \textbf{\bibinfo{volume}{31}},
  \bibinfo{pages}{R1} (\bibinfo{year}{2009}).

\bibitem[{\citenamefont{Rosenbaum et~al.}(2016)\citenamefont{Rosenbaum,
  Kachenoura, Koch, Paques, Cluzel, Redheuil, and Girerd}}]{Rosenbaum2016}
\bibinfo{author}{\bibfnamefont{D.}~\bibnamefont{Rosenbaum}},
  \bibinfo{author}{\bibfnamefont{N.}~\bibnamefont{Kachenoura}},
  \bibinfo{author}{\bibfnamefont{E.}~\bibnamefont{Koch}},
  \bibinfo{author}{\bibfnamefont{M.}~\bibnamefont{Paques}},
  \bibinfo{author}{\bibfnamefont{P.}~\bibnamefont{Cluzel}},
  \bibinfo{author}{\bibfnamefont{A.}~\bibnamefont{Redheuil}}, \bibnamefont{and}
  \bibinfo{author}{\bibfnamefont{X.}~\bibnamefont{Girerd}},
  \bibinfo{journal}{Hypertension Research} \textbf{\bibinfo{volume}{39}},
  \bibinfo{pages}{536} (\bibinfo{year}{2016}).

\bibitem[{\citenamefont{Brar and Platt}(1988)}]{Brar1988}
\bibinfo{author}{\bibfnamefont{H.~S.} \bibnamefont{Brar}} \bibnamefont{and}
  \bibinfo{author}{\bibfnamefont{L.~D.} \bibnamefont{Platt}},
  \bibinfo{journal}{American Journal of Obstetrics and Gynecology}
  \textbf{\bibinfo{volume}{159}}, \bibinfo{pages}{559} (\bibinfo{year}{1988}).

\bibitem[{\citenamefont{Nichols and Edwards}(2001)}]{Nichols2001}
\bibinfo{author}{\bibfnamefont{W.~W.} \bibnamefont{Nichols}} \bibnamefont{and}
  \bibinfo{author}{\bibfnamefont{D.~G.} \bibnamefont{Edwards}},
  \bibinfo{journal}{Journal of Cardiovascular Pharmacology and Therapeutics}
  \textbf{\bibinfo{volume}{6}}, \bibinfo{pages}{5} (\bibinfo{year}{2001}).

\bibitem[{\citenamefont{Maulik}(2005)}]{Maulik2005}
\bibinfo{author}{\bibfnamefont{D.}~\bibnamefont{Maulik}}, in
  \emph{\bibinfo{booktitle}{Doppler ultrasound in obstetrics and gynecology}}
  (\bibinfo{publisher}{Springer}, \bibinfo{year}{2005}), pp.
  \bibinfo{pages}{35--56}.

\bibitem[{\citenamefont{Tsuda et~al.}(2014)\citenamefont{Tsuda, Kunikata,
  Shimura, Aizawa, Omodaka, Shiga, Yasuda, Yokoyama, and Nakazawa}}]{Tsuda2014}
\bibinfo{author}{\bibfnamefont{S.}~\bibnamefont{Tsuda}},
  \bibinfo{author}{\bibfnamefont{H.}~\bibnamefont{Kunikata}},
  \bibinfo{author}{\bibfnamefont{M.}~\bibnamefont{Shimura}},
  \bibinfo{author}{\bibfnamefont{N.}~\bibnamefont{Aizawa}},
  \bibinfo{author}{\bibfnamefont{K.}~\bibnamefont{Omodaka}},
  \bibinfo{author}{\bibfnamefont{Y.}~\bibnamefont{Shiga}},
  \bibinfo{author}{\bibfnamefont{M.}~\bibnamefont{Yasuda}},
  \bibinfo{author}{\bibfnamefont{Y.}~\bibnamefont{Yokoyama}}, \bibnamefont{and}
  \bibinfo{author}{\bibfnamefont{T.}~\bibnamefont{Nakazawa}},
  \bibinfo{journal}{Current Eye Research} \textbf{\bibinfo{volume}{39}},
  \bibinfo{pages}{1207} (\bibinfo{year}{2014}).

\bibitem[{\citenamefont{Luft et~al.}(2016)\citenamefont{Luft, Wozniak,
  Aschinger, Fondi, Bata, Werkmeister, Schmidl, Witkowska, Bolz, Garh{\"o}fer
  et~al.}}]{Luft2016}
\bibinfo{author}{\bibfnamefont{N.}~\bibnamefont{Luft}},
  \bibinfo{author}{\bibfnamefont{P.~A.} \bibnamefont{Wozniak}},
  \bibinfo{author}{\bibfnamefont{G.~C.} \bibnamefont{Aschinger}},
  \bibinfo{author}{\bibfnamefont{K.}~\bibnamefont{Fondi}},
  \bibinfo{author}{\bibfnamefont{A.~M.} \bibnamefont{Bata}},
  \bibinfo{author}{\bibfnamefont{R.~M.} \bibnamefont{Werkmeister}},
  \bibinfo{author}{\bibfnamefont{D.}~\bibnamefont{Schmidl}},
  \bibinfo{author}{\bibfnamefont{K.~J.} \bibnamefont{Witkowska}},
  \bibinfo{author}{\bibfnamefont{M.}~\bibnamefont{Bolz}},
  \bibinfo{author}{\bibfnamefont{G.}~\bibnamefont{Garh{\"o}fer}},
  \bibnamefont{et~al.}, \bibinfo{journal}{PLoS One}
  \textbf{\bibinfo{volume}{11}}, \bibinfo{pages}{e0168190}
  (\bibinfo{year}{2016}).

\bibitem[{\citenamefont{Gu et~al.}(2018)\citenamefont{Gu, Wang, Twa, Tam,
  Girkin, and Zhang}}]{GuZhang2018}
\bibinfo{author}{\bibfnamefont{B.}~\bibnamefont{Gu}},
  \bibinfo{author}{\bibfnamefont{X.}~\bibnamefont{Wang}},
  \bibinfo{author}{\bibfnamefont{M.~D.} \bibnamefont{Twa}},
  \bibinfo{author}{\bibfnamefont{J.}~\bibnamefont{Tam}},
  \bibinfo{author}{\bibfnamefont{C.~A.} \bibnamefont{Girkin}},
  \bibnamefont{and} \bibinfo{author}{\bibfnamefont{Y.}~\bibnamefont{Zhang}},
  \bibinfo{journal}{Biomedical Optics Express} \textbf{\bibinfo{volume}{9}},
  \bibinfo{pages}{3653} (\bibinfo{year}{2018}).

\bibitem[{\citenamefont{Plumb et~al.}(2011)\citenamefont{Plumb, Hamilton, Rea,
  Wright, Hughes, McGivern, and McVeigh}}]{Plumb2011}
\bibinfo{author}{\bibfnamefont{R.~D.} \bibnamefont{Plumb}},
  \bibinfo{author}{\bibfnamefont{P.~K.} \bibnamefont{Hamilton}},
  \bibinfo{author}{\bibfnamefont{D.~J.} \bibnamefont{Rea}},
  \bibinfo{author}{\bibfnamefont{S.~A.} \bibnamefont{Wright}},
  \bibinfo{author}{\bibfnamefont{S.~M.} \bibnamefont{Hughes}},
  \bibinfo{author}{\bibfnamefont{R.~C.} \bibnamefont{McGivern}},
  \bibnamefont{and} \bibinfo{author}{\bibfnamefont{G.~E.}
  \bibnamefont{McVeigh}}, \bibinfo{journal}{The British Journal of Diabetes \&
  Vascular Disease} \textbf{\bibinfo{volume}{11}}, \bibinfo{pages}{243}
  (\bibinfo{year}{2011}).

\bibitem[{\citenamefont{McVeigh et~al.}(2002)\citenamefont{McVeigh, Hamilton,
  and Morgan}}]{Mcveigh2002}
\bibinfo{author}{\bibfnamefont{G.~E.} \bibnamefont{McVeigh}},
  \bibinfo{author}{\bibfnamefont{P.~K.} \bibnamefont{Hamilton}},
  \bibnamefont{and} \bibinfo{author}{\bibfnamefont{D.~R.}
  \bibnamefont{Morgan}}, \bibinfo{journal}{Clinical Science}
  \textbf{\bibinfo{volume}{102}}, \bibinfo{pages}{51} (\bibinfo{year}{2002}).

\bibitem[{\citenamefont{Laurent et~al.}(2006)\citenamefont{Laurent, Cockcroft,
  Van~Bortel, Boutouyrie, Giannattasio, Hayoz, Pannier, Vlachopoulos,
  Wilkinson, and Struijker-Boudier}}]{Laurent2006}
\bibinfo{author}{\bibfnamefont{S.}~\bibnamefont{Laurent}},
  \bibinfo{author}{\bibfnamefont{J.}~\bibnamefont{Cockcroft}},
  \bibinfo{author}{\bibfnamefont{L.}~\bibnamefont{Van~Bortel}},
  \bibinfo{author}{\bibfnamefont{P.}~\bibnamefont{Boutouyrie}},
  \bibinfo{author}{\bibfnamefont{C.}~\bibnamefont{Giannattasio}},
  \bibinfo{author}{\bibfnamefont{D.}~\bibnamefont{Hayoz}},
  \bibinfo{author}{\bibfnamefont{B.}~\bibnamefont{Pannier}},
  \bibinfo{author}{\bibfnamefont{C.}~\bibnamefont{Vlachopoulos}},
  \bibinfo{author}{\bibfnamefont{I.}~\bibnamefont{Wilkinson}},
  \bibnamefont{and}
  \bibinfo{author}{\bibfnamefont{H.}~\bibnamefont{Struijker-Boudier}},
  \bibinfo{journal}{European Heart Journal} \textbf{\bibinfo{volume}{27}},
  \bibinfo{pages}{2588} (\bibinfo{year}{2006}).

\bibitem[{\citenamefont{Safar}(2018)}]{Safar2018}
\bibinfo{author}{\bibfnamefont{M.~E.} \bibnamefont{Safar}},
  \bibinfo{journal}{Nature Reviews Cardiology} \textbf{\bibinfo{volume}{15}},
  \bibinfo{pages}{97} (\bibinfo{year}{2018}).

\bibitem[{\citenamefont{Safar and Lacolley}(2007)}]{Safar2007}
\bibinfo{author}{\bibfnamefont{M.~E.} \bibnamefont{Safar}} \bibnamefont{and}
  \bibinfo{author}{\bibfnamefont{P.}~\bibnamefont{Lacolley}},
  \bibinfo{journal}{American Journal of Physiology-Heart and Circulatory
  Physiology} \textbf{\bibinfo{volume}{293}}, \bibinfo{pages}{H1}
  (\bibinfo{year}{2007}).

\bibitem[{\citenamefont{Bonner and Nossal}(1990)}]{Bonner1990}
\bibinfo{author}{\bibfnamefont{R.~F.} \bibnamefont{Bonner}} \bibnamefont{and}
  \bibinfo{author}{\bibfnamefont{R.}~\bibnamefont{Nossal}}, in
  \emph{\bibinfo{booktitle}{Laser-Doppler blood flowmetry}}
  (\bibinfo{publisher}{Springer}, \bibinfo{year}{1990}), pp.
  \bibinfo{pages}{17--45}.

\bibitem[{\citenamefont{Riva et~al.}(1994)\citenamefont{Riva, Cranstoun,
  Grunwald, and Petrig}}]{Riva1994}
\bibinfo{author}{\bibfnamefont{C.}~\bibnamefont{Riva}},
  \bibinfo{author}{\bibfnamefont{S.}~\bibnamefont{Cranstoun}},
  \bibinfo{author}{\bibfnamefont{J.}~\bibnamefont{Grunwald}}, \bibnamefont{and}
  \bibinfo{author}{\bibfnamefont{B.}~\bibnamefont{Petrig}},
  \bibinfo{journal}{Invest. Ophthalmol. Vis. Sci.}
  \textbf{\bibinfo{volume}{35}}, \bibinfo{pages}{4273} (\bibinfo{year}{1994}).

\bibitem[{\citenamefont{Riva et~al.}(2010)\citenamefont{Riva, Geiser, and
  Petrig}}]{Riva2010}
\bibinfo{author}{\bibfnamefont{C.~E.} \bibnamefont{Riva}},
  \bibinfo{author}{\bibfnamefont{M.}~\bibnamefont{Geiser}}, \bibnamefont{and}
  \bibinfo{author}{\bibfnamefont{B.~L.} \bibnamefont{Petrig}},
  \bibinfo{journal}{Acta Ophthalmologica} \textbf{\bibinfo{volume}{88}},
  \bibinfo{pages}{622} (\bibinfo{year}{2010}).

\bibitem[{\citenamefont{Serov and Lasser}(2005)}]{SerovLasser2005}
\bibinfo{author}{\bibfnamefont{A.}~\bibnamefont{Serov}} \bibnamefont{and}
  \bibinfo{author}{\bibfnamefont{T.}~\bibnamefont{Lasser}},
  \bibinfo{journal}{Optics Express} \textbf{\bibinfo{volume}{13}},
  \bibinfo{pages}{6416} (\bibinfo{year}{2005}).

\bibitem[{\citenamefont{Michelson et~al.}(1996)\citenamefont{Michelson,
  Schmauss, Langhans, Harazny, and Groh}}]{Michelson1996}
\bibinfo{author}{\bibfnamefont{G.}~\bibnamefont{Michelson}},
  \bibinfo{author}{\bibfnamefont{B.}~\bibnamefont{Schmauss}},
  \bibinfo{author}{\bibfnamefont{M.}~\bibnamefont{Langhans}},
  \bibinfo{author}{\bibfnamefont{J.}~\bibnamefont{Harazny}}, \bibnamefont{and}
  \bibinfo{author}{\bibfnamefont{M.}~\bibnamefont{Groh}}, \bibinfo{journal}{J.
  Glaucoma.} \textbf{\bibinfo{volume}{5}}, \bibinfo{pages}{99}
  (\bibinfo{year}{1996}).

\bibitem[{\citenamefont{Mujat et~al.}(2019)\citenamefont{Mujat, Lu, Maguluri,
  Zhao, Iftimia, and Ferguson}}]{Mujat2019}
\bibinfo{author}{\bibfnamefont{M.}~\bibnamefont{Mujat}},
  \bibinfo{author}{\bibfnamefont{Y.}~\bibnamefont{Lu}},
  \bibinfo{author}{\bibfnamefont{G.}~\bibnamefont{Maguluri}},
  \bibinfo{author}{\bibfnamefont{Y.}~\bibnamefont{Zhao}},
  \bibinfo{author}{\bibfnamefont{N.}~\bibnamefont{Iftimia}}, \bibnamefont{and}
  \bibinfo{author}{\bibfnamefont{R.~D.} \bibnamefont{Ferguson}},
  \bibinfo{journal}{Biomedical Optics Express} \textbf{\bibinfo{volume}{10}},
  \bibinfo{pages}{167} (\bibinfo{year}{2019}).

\bibitem[{\citenamefont{Magnain et~al.}(2014)\citenamefont{Magnain, Castel,
  Boucneau, Simonutti, Ferezou, Rancillac, Vitalis, Sahel, Paques, and
  Atlan}}]{MagnainCastelBoucneau2014}
\bibinfo{author}{\bibfnamefont{C.}~\bibnamefont{Magnain}},
  \bibinfo{author}{\bibfnamefont{A.}~\bibnamefont{Castel}},
  \bibinfo{author}{\bibfnamefont{T.}~\bibnamefont{Boucneau}},
  \bibinfo{author}{\bibfnamefont{M.}~\bibnamefont{Simonutti}},
  \bibinfo{author}{\bibfnamefont{I.}~\bibnamefont{Ferezou}},
  \bibinfo{author}{\bibfnamefont{A.}~\bibnamefont{Rancillac}},
  \bibinfo{author}{\bibfnamefont{T.}~\bibnamefont{Vitalis}},
  \bibinfo{author}{\bibfnamefont{J.-A.} \bibnamefont{Sahel}},
  \bibinfo{author}{\bibfnamefont{M.}~\bibnamefont{Paques}}, \bibnamefont{and}
  \bibinfo{author}{\bibfnamefont{M.}~\bibnamefont{Atlan}},
  \bibinfo{journal}{JOSA A} \textbf{\bibinfo{volume}{31}},
  \bibinfo{pages}{2723} (\bibinfo{year}{2014}).

\bibitem[{\citenamefont{Pellizzari et~al.}(2016)\citenamefont{Pellizzari,
  Simonutti, Degardin, Sahel, Fink, Paques, and Atlan}}]{Pellizzari2016}
\bibinfo{author}{\bibfnamefont{M.}~\bibnamefont{Pellizzari}},
  \bibinfo{author}{\bibfnamefont{M.}~\bibnamefont{Simonutti}},
  \bibinfo{author}{\bibfnamefont{J.}~\bibnamefont{Degardin}},
  \bibinfo{author}{\bibfnamefont{J.-A.} \bibnamefont{Sahel}},
  \bibinfo{author}{\bibfnamefont{M.}~\bibnamefont{Fink}},
  \bibinfo{author}{\bibfnamefont{M.}~\bibnamefont{Paques}}, \bibnamefont{and}
  \bibinfo{author}{\bibfnamefont{M.}~\bibnamefont{Atlan}},
  \bibinfo{journal}{Optics Letters} \textbf{\bibinfo{volume}{41}},
  \bibinfo{pages}{3503} (\bibinfo{year}{2016}).

\bibitem[{\citenamefont{Donnarumma et~al.}(2016)\citenamefont{Donnarumma,
  Brodoline, Alexandre, and Gross}}]{Donnarumma2016}
\bibinfo{author}{\bibfnamefont{D.}~\bibnamefont{Donnarumma}},
  \bibinfo{author}{\bibfnamefont{A.}~\bibnamefont{Brodoline}},
  \bibinfo{author}{\bibfnamefont{D.}~\bibnamefont{Alexandre}},
  \bibnamefont{and} \bibinfo{author}{\bibfnamefont{M.}~\bibnamefont{Gross}},
  \bibinfo{journal}{Optics Express} \textbf{\bibinfo{volume}{24}},
  \bibinfo{pages}{26887} (\bibinfo{year}{2016}).

\bibitem[{\citenamefont{Hillmann et~al.}(2016)\citenamefont{Hillmann, Spahr,
  Hain, Sudkamp, Franke, Pf{\"a}ffle, Winter, and H{\"u}ttmann}}]{Hillmann2016}
\bibinfo{author}{\bibfnamefont{D.}~\bibnamefont{Hillmann}},
  \bibinfo{author}{\bibfnamefont{H.}~\bibnamefont{Spahr}},
  \bibinfo{author}{\bibfnamefont{C.}~\bibnamefont{Hain}},
  \bibinfo{author}{\bibfnamefont{H.}~\bibnamefont{Sudkamp}},
  \bibinfo{author}{\bibfnamefont{G.}~\bibnamefont{Franke}},
  \bibinfo{author}{\bibfnamefont{C.}~\bibnamefont{Pf{\"a}ffle}},
  \bibinfo{author}{\bibfnamefont{C.}~\bibnamefont{Winter}}, \bibnamefont{and}
  \bibinfo{author}{\bibfnamefont{G.}~\bibnamefont{H{\"u}ttmann}},
  \bibinfo{journal}{Scientific reports} \textbf{\bibinfo{volume}{6}},
  \bibinfo{pages}{35209} (\bibinfo{year}{2016}).

\bibitem[{\citenamefont{Ginner et~al.}(2018)\citenamefont{Ginner, Schmoll,
  Kumar, Salas, Pricoupenko, Wurster, and Leitgeb}}]{Ginner2018}
\bibinfo{author}{\bibfnamefont{L.}~\bibnamefont{Ginner}},
  \bibinfo{author}{\bibfnamefont{T.}~\bibnamefont{Schmoll}},
  \bibinfo{author}{\bibfnamefont{A.}~\bibnamefont{Kumar}},
  \bibinfo{author}{\bibfnamefont{M.}~\bibnamefont{Salas}},
  \bibinfo{author}{\bibfnamefont{N.}~\bibnamefont{Pricoupenko}},
  \bibinfo{author}{\bibfnamefont{L.~M.} \bibnamefont{Wurster}},
  \bibnamefont{and} \bibinfo{author}{\bibfnamefont{R.~A.}
  \bibnamefont{Leitgeb}}, \bibinfo{journal}{Biomedical Optics Express}
  \textbf{\bibinfo{volume}{9}}, \bibinfo{pages}{472} (\bibinfo{year}{2018}).

\bibitem[{\citenamefont{Spahr et~al.}(2018)\citenamefont{Spahr, Pf{\"a}ffle,
  Koch, Sudkamp, H{\"u}ttmann, and Hillmann}}]{Spahr2018}
\bibinfo{author}{\bibfnamefont{H.}~\bibnamefont{Spahr}},
  \bibinfo{author}{\bibfnamefont{C.}~\bibnamefont{Pf{\"a}ffle}},
  \bibinfo{author}{\bibfnamefont{P.}~\bibnamefont{Koch}},
  \bibinfo{author}{\bibfnamefont{H.}~\bibnamefont{Sudkamp}},
  \bibinfo{author}{\bibfnamefont{G.}~\bibnamefont{H{\"u}ttmann}},
  \bibnamefont{and} \bibinfo{author}{\bibfnamefont{D.}~\bibnamefont{Hillmann}},
  \bibinfo{journal}{Optics express} \textbf{\bibinfo{volume}{26}},
  \bibinfo{pages}{18803} (\bibinfo{year}{2018}).

\bibitem[{\citenamefont{Ginner et~al.}(2019)\citenamefont{Ginner, Wartak,
  Salas, Augustin, Niederleithner, Wurster, and Leitgeb}}]{Ginner2019}
\bibinfo{author}{\bibfnamefont{L.}~\bibnamefont{Ginner}},
  \bibinfo{author}{\bibfnamefont{A.}~\bibnamefont{Wartak}},
  \bibinfo{author}{\bibfnamefont{M.}~\bibnamefont{Salas}},
  \bibinfo{author}{\bibfnamefont{M.}~\bibnamefont{Augustin}},
  \bibinfo{author}{\bibfnamefont{M.}~\bibnamefont{Niederleithner}},
  \bibinfo{author}{\bibfnamefont{L.~M.} \bibnamefont{Wurster}},
  \bibnamefont{and} \bibinfo{author}{\bibfnamefont{R.~A.}
  \bibnamefont{Leitgeb}}, \bibinfo{journal}{Optics Letters}
  \textbf{\bibinfo{volume}{44}}, \bibinfo{pages}{967} (\bibinfo{year}{2019}).

\bibitem[{\citenamefont{Brodoline et~al.}(2019)\citenamefont{Brodoline, Rawat,
  Alexandre, Cubedo, and Gross}}]{Brodoline2019}
\bibinfo{author}{\bibfnamefont{A.}~\bibnamefont{Brodoline}},
  \bibinfo{author}{\bibfnamefont{N.}~\bibnamefont{Rawat}},
  \bibinfo{author}{\bibfnamefont{D.}~\bibnamefont{Alexandre}},
  \bibinfo{author}{\bibfnamefont{N.}~\bibnamefont{Cubedo}}, \bibnamefont{and}
  \bibinfo{author}{\bibfnamefont{M.}~\bibnamefont{Gross}},
  \bibinfo{journal}{Optics Letters} \textbf{\bibinfo{volume}{44}},
  \bibinfo{pages}{2827} (\bibinfo{year}{2019}).

\bibitem[{\citenamefont{Puyo et~al.}(2018)\citenamefont{Puyo, Paques, Fink,
  Sahel, and Atlan}}]{Puyo2018}
\bibinfo{author}{\bibfnamefont{L.}~\bibnamefont{Puyo}},
  \bibinfo{author}{\bibfnamefont{M.}~\bibnamefont{Paques}},
  \bibinfo{author}{\bibfnamefont{M.}~\bibnamefont{Fink}},
  \bibinfo{author}{\bibfnamefont{J.-A.} \bibnamefont{Sahel}}, \bibnamefont{and}
  \bibinfo{author}{\bibfnamefont{M.}~\bibnamefont{Atlan}},
  \bibinfo{journal}{Biomed. Opt. Express} \textbf{\bibinfo{volume}{9}},
  \bibinfo{pages}{4113} (\bibinfo{year}{2018}).

\bibitem[{\citenamefont{Puyo et~al.}(2019)\citenamefont{Puyo, Paques, Fink,
  Sahel, and Atlan}}]{Puyo2019}
\bibinfo{author}{\bibfnamefont{L.}~\bibnamefont{Puyo}},
  \bibinfo{author}{\bibfnamefont{M.}~\bibnamefont{Paques}},
  \bibinfo{author}{\bibfnamefont{M.}~\bibnamefont{Fink}},
  \bibinfo{author}{\bibfnamefont{J.-A.} \bibnamefont{Sahel}}, \bibnamefont{and}
  \bibinfo{author}{\bibfnamefont{M.}~\bibnamefont{Atlan}},
  \bibinfo{journal}{Biomedical Optics Express} \textbf{\bibinfo{volume}{10}},
  \bibinfo{pages}{995} (\bibinfo{year}{2019}).

\bibitem[{\citenamefont{Postnov et~al.}(2019)\citenamefont{Postnov, Cheng,
  Erdener, and Boas}}]{Postnov2019}
\bibinfo{author}{\bibfnamefont{D.~D.} \bibnamefont{Postnov}},
  \bibinfo{author}{\bibfnamefont{X.}~\bibnamefont{Cheng}},
  \bibinfo{author}{\bibfnamefont{S.~E.} \bibnamefont{Erdener}},
  \bibnamefont{and} \bibinfo{author}{\bibfnamefont{D.~A.} \bibnamefont{Boas}},
  \bibinfo{journal}{Scientific Reports} \textbf{\bibinfo{volume}{9}},
  \bibinfo{pages}{2542} (\bibinfo{year}{2019}).

\bibitem[{\citenamefont{Goodman}(2005)}]{Goodman2005}
\bibinfo{author}{\bibfnamefont{J.~W.} \bibnamefont{Goodman}},
  \emph{\bibinfo{title}{Introduction to Fourier optics}}
  (\bibinfo{publisher}{Roberts and Company Publishers}, \bibinfo{year}{2005}).

\bibitem[{\citenamefont{Evans}(1989)}]{Evans1989}
\bibinfo{author}{\bibfnamefont{D.~H.} \bibnamefont{Evans}},
  \emph{\bibinfo{title}{Doppler ultrasound: Physics, instrumentation, and
  clinical applications}} (\bibinfo{publisher}{John Wiley \& Sons},
  \bibinfo{year}{1989}).

\bibitem[{\citenamefont{Mac{\'e} et~al.}(2011)\citenamefont{Mac{\'e}, Montaldo,
  Cohen, Baulac, Fink, and Tanter}}]{Mace2011}
\bibinfo{author}{\bibfnamefont{E.}~\bibnamefont{Mac{\'e}}},
  \bibinfo{author}{\bibfnamefont{G.}~\bibnamefont{Montaldo}},
  \bibinfo{author}{\bibfnamefont{I.}~\bibnamefont{Cohen}},
  \bibinfo{author}{\bibfnamefont{M.}~\bibnamefont{Baulac}},
  \bibinfo{author}{\bibfnamefont{M.}~\bibnamefont{Fink}}, \bibnamefont{and}
  \bibinfo{author}{\bibfnamefont{M.}~\bibnamefont{Tanter}},
  \bibinfo{journal}{Nature Methods} \textbf{\bibinfo{volume}{8}},
  \bibinfo{pages}{662} (\bibinfo{year}{2011}).

\bibitem[{\citenamefont{Nichols et~al.}(1980)\citenamefont{Nichols, Pepine,
  Geiser, and Conti}}]{Nichols1980}
\bibinfo{author}{\bibfnamefont{W.}~\bibnamefont{Nichols}},
  \bibinfo{author}{\bibfnamefont{C.}~\bibnamefont{Pepine}},
  \bibinfo{author}{\bibfnamefont{E.}~\bibnamefont{Geiser}}, \bibnamefont{and}
  \bibinfo{author}{\bibfnamefont{C.}~\bibnamefont{Conti}}, in
  \emph{\bibinfo{booktitle}{Federation Proceedings}} (\bibinfo{year}{1980}),
  vol.~\bibinfo{volume}{39}, pp. \bibinfo{pages}{196--201}.

\bibitem[{\citenamefont{Toy et~al.}(1985)\citenamefont{Toy, Melbin, and
  Noordergraaf}}]{Toy1985}
\bibinfo{author}{\bibfnamefont{S.~M.} \bibnamefont{Toy}},
  \bibinfo{author}{\bibfnamefont{J.}~\bibnamefont{Melbin}}, \bibnamefont{and}
  \bibinfo{author}{\bibfnamefont{A.}~\bibnamefont{Noordergraaf}},
  \bibinfo{journal}{IEEE Transactions on Biomedical Engineering} pp.
  \bibinfo{pages}{174--176} (\bibinfo{year}{1985}).

\bibitem[{\citenamefont{Tranquart et~al.}(2003)\citenamefont{Tranquart,
  Berg{\`e}s, Koskas, Arsene, Rossazza, Pisella, and
  Pourcelot}}]{Tranquart2003}
\bibinfo{author}{\bibfnamefont{F.}~\bibnamefont{Tranquart}},
  \bibinfo{author}{\bibfnamefont{O.}~\bibnamefont{Berg{\`e}s}},
  \bibinfo{author}{\bibfnamefont{P.}~\bibnamefont{Koskas}},
  \bibinfo{author}{\bibfnamefont{S.}~\bibnamefont{Arsene}},
  \bibinfo{author}{\bibfnamefont{C.}~\bibnamefont{Rossazza}},
  \bibinfo{author}{\bibfnamefont{P.-J.} \bibnamefont{Pisella}},
  \bibnamefont{and}
  \bibinfo{author}{\bibfnamefont{L.}~\bibnamefont{Pourcelot}},
  \bibinfo{journal}{Journal of Clinical Ultrasound}
  \textbf{\bibinfo{volume}{31}}, \bibinfo{pages}{258} (\bibinfo{year}{2003}).

\bibitem[{\citenamefont{Demen{\'e} et~al.}(2014)\citenamefont{Demen{\'e},
  Pernot, Biran, Alison, Fink, Baud, and Tanter}}]{Demene2014}
\bibinfo{author}{\bibfnamefont{C.}~\bibnamefont{Demen{\'e}}},
  \bibinfo{author}{\bibfnamefont{M.}~\bibnamefont{Pernot}},
  \bibinfo{author}{\bibfnamefont{V.}~\bibnamefont{Biran}},
  \bibinfo{author}{\bibfnamefont{M.}~\bibnamefont{Alison}},
  \bibinfo{author}{\bibfnamefont{M.}~\bibnamefont{Fink}},
  \bibinfo{author}{\bibfnamefont{O.}~\bibnamefont{Baud}}, \bibnamefont{and}
  \bibinfo{author}{\bibfnamefont{M.}~\bibnamefont{Tanter}},
  \bibinfo{journal}{Journal of Cerebral Blood Flow \& Metabolism}
  \textbf{\bibinfo{volume}{34}}, \bibinfo{pages}{1009} (\bibinfo{year}{2014}).

\bibitem[{\citenamefont{Pourcelot}(1974)}]{Pourcelot1974}
\bibinfo{author}{\bibfnamefont{L.}~\bibnamefont{Pourcelot}},
  \bibinfo{journal}{Velocimetrie Ultrasonore Doppler}
  \textbf{\bibinfo{volume}{34}}, \bibinfo{pages}{780} (\bibinfo{year}{1974}).

\bibitem[{\citenamefont{Bude and
  Rubin}(1999{\natexlab{a}})}]{Bude1999Relationship}
\bibinfo{author}{\bibfnamefont{R.~O.} \bibnamefont{Bude}} \bibnamefont{and}
  \bibinfo{author}{\bibfnamefont{J.~M.} \bibnamefont{Rubin}},
  \bibinfo{journal}{Radiology} \textbf{\bibinfo{volume}{211}},
  \bibinfo{pages}{411} (\bibinfo{year}{1999}{\natexlab{a}}).

\bibitem[{\citenamefont{Bude and
  Rubin}(1999{\natexlab{b}})}]{Bude1999Downstream}
\bibinfo{author}{\bibfnamefont{R.~O.} \bibnamefont{Bude}} \bibnamefont{and}
  \bibinfo{author}{\bibfnamefont{J.~M.} \bibnamefont{Rubin}},
  \bibinfo{journal}{Radiology} \textbf{\bibinfo{volume}{212}},
  \bibinfo{pages}{732} (\bibinfo{year}{1999}{\natexlab{b}}).

\bibitem[{\citenamefont{Halpern et~al.}(1998)\citenamefont{Halpern, Merton, and
  Forsberg}}]{Halpern1998}
\bibinfo{author}{\bibfnamefont{E.~J.} \bibnamefont{Halpern}},
  \bibinfo{author}{\bibfnamefont{D.~A.} \bibnamefont{Merton}},
  \bibnamefont{and} \bibinfo{author}{\bibfnamefont{F.}~\bibnamefont{Forsberg}},
  \bibinfo{journal}{Radiology} \textbf{\bibinfo{volume}{206}},
  \bibinfo{pages}{761} (\bibinfo{year}{1998}).

\bibitem[{\citenamefont{Polska et~al.}(2001)\citenamefont{Polska, Kircher,
  Ehrlich, Vecsei, and Schmetterer}}]{Polska2001}
\bibinfo{author}{\bibfnamefont{E.}~\bibnamefont{Polska}},
  \bibinfo{author}{\bibfnamefont{K.}~\bibnamefont{Kircher}},
  \bibinfo{author}{\bibfnamefont{P.}~\bibnamefont{Ehrlich}},
  \bibinfo{author}{\bibfnamefont{P.~V.} \bibnamefont{Vecsei}},
  \bibnamefont{and}
  \bibinfo{author}{\bibfnamefont{L.}~\bibnamefont{Schmetterer}},
  \bibinfo{journal}{American Journal of Physiology-Heart and Circulatory
  Physiology} \textbf{\bibinfo{volume}{280}}, \bibinfo{pages}{H1442}
  (\bibinfo{year}{2001}).

\bibitem[{\citenamefont{Network}(2011)}]{Anand2010}
\bibinfo{author}{\bibfnamefont{C.~V.} \bibnamefont{Network}},
  \bibinfo{journal}{The Retina and Its Disorders}
  \textbf{\bibinfo{volume}{179}} (\bibinfo{year}{2011}).

\bibitem[{\citenamefont{Isono et~al.}(2003)\citenamefont{Isono, Kishi, Kimura,
  Hagiwara, Konishi, and Fujii}}]{Isono2003}
\bibinfo{author}{\bibfnamefont{H.}~\bibnamefont{Isono}},
  \bibinfo{author}{\bibfnamefont{S.}~\bibnamefont{Kishi}},
  \bibinfo{author}{\bibfnamefont{Y.}~\bibnamefont{Kimura}},
  \bibinfo{author}{\bibfnamefont{N.}~\bibnamefont{Hagiwara}},
  \bibinfo{author}{\bibfnamefont{N.}~\bibnamefont{Konishi}}, \bibnamefont{and}
  \bibinfo{author}{\bibfnamefont{H.}~\bibnamefont{Fujii}},
  \bibinfo{journal}{Archives of Ophthalmology} \textbf{\bibinfo{volume}{121}},
  \bibinfo{pages}{225} (\bibinfo{year}{2003}).

\bibitem[{\citenamefont{Iwase et~al.}(2015)\citenamefont{Iwase, Yamamoto, Ra,
  Murotani, Matsui, and Terasaki}}]{Iwase2015}
\bibinfo{author}{\bibfnamefont{T.}~\bibnamefont{Iwase}},
  \bibinfo{author}{\bibfnamefont{K.}~\bibnamefont{Yamamoto}},
  \bibinfo{author}{\bibfnamefont{E.}~\bibnamefont{Ra}},
  \bibinfo{author}{\bibfnamefont{K.}~\bibnamefont{Murotani}},
  \bibinfo{author}{\bibfnamefont{S.}~\bibnamefont{Matsui}}, \bibnamefont{and}
  \bibinfo{author}{\bibfnamefont{H.}~\bibnamefont{Terasaki}},
  \bibinfo{journal}{Medicine} \textbf{\bibinfo{volume}{94}}
  (\bibinfo{year}{2015}).

\bibitem[{\citenamefont{Fondi et~al.}(2018)\citenamefont{Fondi, Bata, Luft,
  Witkowska, Werkmeister, Schmidl, Bolz, Schmetterer, and
  Garh{\"o}fer}}]{Fondi2018}
\bibinfo{author}{\bibfnamefont{K.}~\bibnamefont{Fondi}},
  \bibinfo{author}{\bibfnamefont{A.~M.} \bibnamefont{Bata}},
  \bibinfo{author}{\bibfnamefont{N.}~\bibnamefont{Luft}},
  \bibinfo{author}{\bibfnamefont{K.~J.} \bibnamefont{Witkowska}},
  \bibinfo{author}{\bibfnamefont{R.~M.} \bibnamefont{Werkmeister}},
  \bibinfo{author}{\bibfnamefont{D.}~\bibnamefont{Schmidl}},
  \bibinfo{author}{\bibfnamefont{M.}~\bibnamefont{Bolz}},
  \bibinfo{author}{\bibfnamefont{L.}~\bibnamefont{Schmetterer}},
  \bibnamefont{and}
  \bibinfo{author}{\bibfnamefont{G.}~\bibnamefont{Garh{\"o}fer}},
  \bibinfo{journal}{PloS One} \textbf{\bibinfo{volume}{13}},
  \bibinfo{pages}{e0207525} (\bibinfo{year}{2018}).

\bibitem[{\citenamefont{Squirrell et~al.}(2001)\citenamefont{Squirrell, Watts,
  Evans, Mody, and Talbot}}]{Squirrell2001}
\bibinfo{author}{\bibfnamefont{D.}~\bibnamefont{Squirrell}},
  \bibinfo{author}{\bibfnamefont{A.}~\bibnamefont{Watts}},
  \bibinfo{author}{\bibfnamefont{D.}~\bibnamefont{Evans}},
  \bibinfo{author}{\bibfnamefont{C.}~\bibnamefont{Mody}}, \bibnamefont{and}
  \bibinfo{author}{\bibfnamefont{J.}~\bibnamefont{Talbot}},
  \bibinfo{journal}{Eye} \textbf{\bibinfo{volume}{15}}, \bibinfo{pages}{261}
  (\bibinfo{year}{2001}).

\bibitem[{\citenamefont{Lieb et~al.}(1991)\citenamefont{Lieb, Cohen, Merton,
  Shields, Mitchell, and Goldberg}}]{Lieb1991}
\bibinfo{author}{\bibfnamefont{W.~E.} \bibnamefont{Lieb}},
  \bibinfo{author}{\bibfnamefont{S.~M.} \bibnamefont{Cohen}},
  \bibinfo{author}{\bibfnamefont{D.~A.} \bibnamefont{Merton}},
  \bibinfo{author}{\bibfnamefont{J.~A.} \bibnamefont{Shields}},
  \bibinfo{author}{\bibfnamefont{D.~G.} \bibnamefont{Mitchell}},
  \bibnamefont{and} \bibinfo{author}{\bibfnamefont{B.~B.}
  \bibnamefont{Goldberg}}, \bibinfo{journal}{Archives of Ophthalmology}
  \textbf{\bibinfo{volume}{109}}, \bibinfo{pages}{527} (\bibinfo{year}{1991}).

\bibitem[{\citenamefont{Morgan et~al.}(2016)\citenamefont{Morgan, Hazelton, and
  Yu}}]{Morgan2016}
\bibinfo{author}{\bibfnamefont{W.~H.} \bibnamefont{Morgan}},
  \bibinfo{author}{\bibfnamefont{M.~L.} \bibnamefont{Hazelton}},
  \bibnamefont{and} \bibinfo{author}{\bibfnamefont{D.-Y.} \bibnamefont{Yu}},
  \bibinfo{journal}{Progress In Retinal And Eye Research}
  \textbf{\bibinfo{volume}{55}}, \bibinfo{pages}{82} (\bibinfo{year}{2016}).

\bibitem[{\citenamefont{Wartak et~al.}(2019)\citenamefont{Wartak, Beer,
  Desissaire, Baumann, Pircher, and Hitzenberger}}]{Wartak2019}
\bibinfo{author}{\bibfnamefont{A.}~\bibnamefont{Wartak}},
  \bibinfo{author}{\bibfnamefont{F.}~\bibnamefont{Beer}},
  \bibinfo{author}{\bibfnamefont{S.}~\bibnamefont{Desissaire}},
  \bibinfo{author}{\bibfnamefont{B.}~\bibnamefont{Baumann}},
  \bibinfo{author}{\bibfnamefont{M.}~\bibnamefont{Pircher}}, \bibnamefont{and}
  \bibinfo{author}{\bibfnamefont{C.~K.} \bibnamefont{Hitzenberger}},
  \bibinfo{journal}{Scientific Reports} \textbf{\bibinfo{volume}{9}},
  \bibinfo{pages}{4237} (\bibinfo{year}{2019}).

\bibitem[{\citenamefont{Kain et~al.}(2010)\citenamefont{Kain, Morgan, and
  Yu}}]{Kain2010}
\bibinfo{author}{\bibfnamefont{S.}~\bibnamefont{Kain}},
  \bibinfo{author}{\bibfnamefont{W.~H.} \bibnamefont{Morgan}},
  \bibnamefont{and} \bibinfo{author}{\bibfnamefont{D.-Y.} \bibnamefont{Yu}},
  \bibinfo{journal}{British Journal of Ophthalmology}
  \textbf{\bibinfo{volume}{94}}, \bibinfo{pages}{854} (\bibinfo{year}{2010}).

\bibitem[{\citenamefont{Morgan et~al.}(2012)\citenamefont{Morgan, Lind, Kain,
  Fatehee, Bala, and Yu}}]{Morgan2012}
\bibinfo{author}{\bibfnamefont{W.~H.} \bibnamefont{Morgan}},
  \bibinfo{author}{\bibfnamefont{C.~R.} \bibnamefont{Lind}},
  \bibinfo{author}{\bibfnamefont{S.}~\bibnamefont{Kain}},
  \bibinfo{author}{\bibfnamefont{N.}~\bibnamefont{Fatehee}},
  \bibinfo{author}{\bibfnamefont{A.}~\bibnamefont{Bala}}, \bibnamefont{and}
  \bibinfo{author}{\bibfnamefont{D.-Y.} \bibnamefont{Yu}},
  \bibinfo{journal}{Investigative Ophthalmology \& Visual Science}
  \textbf{\bibinfo{volume}{53}}, \bibinfo{pages}{4676} (\bibinfo{year}{2012}).

\bibitem[{\citenamefont{Morgan et~al.}(2014)\citenamefont{Morgan, Hazelton,
  Betz-Stablein, Yu, Lind, Ravichandran, and House}}]{Morgan2014}
\bibinfo{author}{\bibfnamefont{W.~H.} \bibnamefont{Morgan}},
  \bibinfo{author}{\bibfnamefont{M.~L.} \bibnamefont{Hazelton}},
  \bibinfo{author}{\bibfnamefont{B.~D.} \bibnamefont{Betz-Stablein}},
  \bibinfo{author}{\bibfnamefont{D.-Y.} \bibnamefont{Yu}},
  \bibinfo{author}{\bibfnamefont{C.~R.} \bibnamefont{Lind}},
  \bibinfo{author}{\bibfnamefont{V.}~\bibnamefont{Ravichandran}},
  \bibnamefont{and} \bibinfo{author}{\bibfnamefont{P.~H.} \bibnamefont{House}},
  \bibinfo{journal}{Investigative Ophthalmology \& Visual Science}
  \textbf{\bibinfo{volume}{55}}, \bibinfo{pages}{5998} (\bibinfo{year}{2014}).

\bibitem[{\citenamefont{Moret et~al.}(2015)\citenamefont{Moret, Reiff, Lagreze,
  and Bach}}]{Moret2015}
\bibinfo{author}{\bibfnamefont{F.}~\bibnamefont{Moret}},
  \bibinfo{author}{\bibfnamefont{C.~M.} \bibnamefont{Reiff}},
  \bibinfo{author}{\bibfnamefont{W.~A.} \bibnamefont{Lagreze}},
  \bibnamefont{and} \bibinfo{author}{\bibfnamefont{M.}~\bibnamefont{Bach}},
  \bibinfo{journal}{Translational Vision Science \& Technology}
  \textbf{\bibinfo{volume}{4}}, \bibinfo{pages}{3} (\bibinfo{year}{2015}).

\bibitem[{\citenamefont{Tan et~al.}(2015)\citenamefont{Tan, Liu, Liang, Gao,
  Pechauer, Jia, and Huang}}]{Tan2015}
\bibinfo{author}{\bibfnamefont{O.}~\bibnamefont{Tan}},
  \bibinfo{author}{\bibfnamefont{G.}~\bibnamefont{Liu}},
  \bibinfo{author}{\bibfnamefont{L.}~\bibnamefont{Liang}},
  \bibinfo{author}{\bibfnamefont{S.~S.} \bibnamefont{Gao}},
  \bibinfo{author}{\bibfnamefont{A.~D.} \bibnamefont{Pechauer}},
  \bibinfo{author}{\bibfnamefont{Y.}~\bibnamefont{Jia}}, \bibnamefont{and}
  \bibinfo{author}{\bibfnamefont{D.}~\bibnamefont{Huang}},
  \bibinfo{journal}{Journal Of Biomedical Optics}
  \textbf{\bibinfo{volume}{20}}, \bibinfo{pages}{066004}
  (\bibinfo{year}{2015}).

\bibitem[{\citenamefont{Doblhoff-Dier et~al.}(2014)\citenamefont{Doblhoff-Dier,
  Schmetterer, Vilser, Garh{\"o}fer, Gr{\"o}schl, Leitgeb, and
  Werkmeister}}]{Doblhoff2014}
\bibinfo{author}{\bibfnamefont{V.}~\bibnamefont{Doblhoff-Dier}},
  \bibinfo{author}{\bibfnamefont{L.}~\bibnamefont{Schmetterer}},
  \bibinfo{author}{\bibfnamefont{W.}~\bibnamefont{Vilser}},
  \bibinfo{author}{\bibfnamefont{G.}~\bibnamefont{Garh{\"o}fer}},
  \bibinfo{author}{\bibfnamefont{M.}~\bibnamefont{Gr{\"o}schl}},
  \bibinfo{author}{\bibfnamefont{R.~A.} \bibnamefont{Leitgeb}},
  \bibnamefont{and} \bibinfo{author}{\bibfnamefont{R.~M.}
  \bibnamefont{Werkmeister}}, \bibinfo{journal}{Biomedical Optics Express}
  \textbf{\bibinfo{volume}{5}}, \bibinfo{pages}{630} (\bibinfo{year}{2014}).

\bibitem[{\citenamefont{Michelson and Harazny}(1997)}]{Michelson1997}
\bibinfo{author}{\bibfnamefont{G.}~\bibnamefont{Michelson}} \bibnamefont{and}
  \bibinfo{author}{\bibfnamefont{J.}~\bibnamefont{Harazny}},
  \bibinfo{journal}{Ophthalmology} \textbf{\bibinfo{volume}{104}},
  \bibinfo{pages}{664} (\bibinfo{year}{1997}).

\bibitem[{\citenamefont{Paques et~al.}(2005)\citenamefont{Paques, Baillart,
  Genevois, Gaudric, L{\'e}vy, and Sahel}}]{Paques2005}
\bibinfo{author}{\bibfnamefont{M.}~\bibnamefont{Paques}},
  \bibinfo{author}{\bibfnamefont{O.}~\bibnamefont{Baillart}},
  \bibinfo{author}{\bibfnamefont{O.}~\bibnamefont{Genevois}},
  \bibinfo{author}{\bibfnamefont{A.}~\bibnamefont{Gaudric}},
  \bibinfo{author}{\bibfnamefont{B.}~\bibnamefont{L{\'e}vy}}, \bibnamefont{and}
  \bibinfo{author}{\bibfnamefont{J.}~\bibnamefont{Sahel}},
  \bibinfo{journal}{British Journal of Ophthalmology}
  \textbf{\bibinfo{volume}{89}}, \bibinfo{pages}{1036} (\bibinfo{year}{2005}).

\end{thebibliography}

\end{document}